\documentclass[journal,twoside,web]{ieeecolor}
\usepackage{tmi}
\usepackage{cite}
\usepackage{url}
\usepackage{amsmath,amssymb,amsfonts}
\usepackage{algorithmic}
\usepackage{mathtools} 
\usepackage{multirow}
\usepackage{graphicx}
\usepackage{overpic}
\usepackage{textcomp}
\usepackage{floatrow}
\usepackage{arydshln} 
\usepackage{url}
\usepackage{caption}
\usepackage{xcolor}
\usepackage{bbding}
\usepackage{hyperref}
\usepackage{graphicx}
\usepackage[misc]{ifsym}
\usepackage{pgfplots}
\usepackage{tikz}
\usepackage[ruled]{algorithm2e}
\usepackage{multirow}
\usepackage{xr}
\usepackage{soul} 
\usepackage{graphicx} 
\usepackage{float} 
\usepackage{subfigure}  
\usepackage{sidecap}
\usepackage{scrextend}
\usepackage[switch]{lineno}

\newcommand{\hly}[1]{{#1}}

\newfloatcommand{capbtabbox}{table}[][\FBwidth]

\SetCommentSty{mycommfont}
\usepackage{bm}
\newcommand{\etal}{\textit{et al. }}
\newcommand{\ie}{i.e., }
\newcommand{\MyFig}{Fig. }

\newcommand{\vs}{vs. }

\makeatletter
\newcommand*{\addFileDependency}[1]{
	\typeout{(#1)}
	\@addtofilelist{#1}
	\IfFileExists{#1}{}{\typeout{No file #1.}}
}
\makeatother

\newcommand*{\myexternaldocument}[1]{
	\externaldocument{#1}
	\addFileDependency{#1.tex}
	\addFileDependency{#1.aux}
}

\myexternaldocument{supply}
\definecolor{newcolor}{rgb}{.8,.349,.1}

\def\BibTeX{{\rm B\kern-.05em{\sc i\kern-.025em b}\kern-.08em
		T\kern-.1667em\lower.7ex\hbox{E}\kern-.125emX}}
\begin{document}
	
	\title{
		Aligning Multi-Sequence CMR Towards Fully Automated Myocardial Pathology Segmentation
	}
	
	\author{
		Wangbin Ding, Lei Li, Junyi Qiu, Sihan Wang,  Liqin Huang, Yinyin Chen, Shan Yang,  Xiahai Zhuang
		\thanks{L Huang and X Zhuang are co-senior authors and contribute equally. Corresponding authors: L Huang, X Zhuang, Y Chen and S Yang.  This work was supported by the National Nature Science Foundation of China, 62111530195, 61971142 and 62271149; Fujian Provincial Natural Science Foundation Project 2021J02019. The work of L. Li was supported by the SJTU 2021 Outstanding Doctoral Graduate Development Scholarship.} 
		\thanks{W Ding and L Huang are with the College of Physics and Information Engineering, Fuzhou University, Fuzhou 350117, China (n191110003@fzu.edu.cn; hlq@fzu.edu.cn).} 
		\thanks{L Li is with the Department of Engineering Science, University of Oxford, Oxford, UK (e-mail: lei.li@eng.ox.ac.uk).}
		\thanks{J Qiu, S Wang and X Zhuang are with the School of Data Science, Fudan University, Shanghai, China (qjy980811@gmail.com; 21110980009@m.fudan.edu.cn; zxh@fudan.edu.cn).}
		\thanks{Y Chen and S Yang are with the Department of Radiology, Zhongshan Hospital, Fudan University and also the Department of Medical Imaging, Shanghai Medical School, Fudan University and Shanghai Institute of Medical Imaging, Shanghai 200032, China (chen.yinyin@zs-hospital.sh.cn; yang.shan@zs-hospital.sh.cn).}
	}

	\maketitle
	
	\begin{abstract}
		Myocardial pathology segmentation (MyoPS) is critical for the risk stratification and treatment planning of myocardial infarction (MI). Multi-sequence cardiac magnetic resonance (MS-CMR) images  can provide valuable information. For instance, balanced steady-state free precession  cine sequences present clear anatomical boundaries, while late gadolinium enhancement and T2-weighted CMR sequences visualize myocardial scar and edema of MI, respectively. Existing methods usually fuse anatomical and pathological information from different CMR sequences for MyoPS, but assume that these images have been spatially aligned. However, \hly{MS-CMR} images are usually unaligned due to the respiratory motions in clinical practices, which poses additional challenges for MyoPS. This work presents an automatic MyoPS framework for unaligned \hly{MS-CMR} images. Specifically, we design a combined computing model for simultaneous image registration and information fusion, which aggregates multi-sequence features into a common space to extract anatomical structures (\ie myocardium). Consequently, we can highlight the informative regions in the common space via the extracted myocardium to improve MyoPS performance, considering the spatial relationship between myocardial pathologies and myocardium. Experiments on a private \hly{MS-CMR} dataset and a public dataset from the MYOPS2020 challenge show that our framework could achieve promising performance for fully automatic MyoPS.
	\end{abstract}
	
	\begin{IEEEkeywords}
		Myocardial pathology, Multi-sequence cardiac magnetic resonance,  Registration, Segmentation
	\end{IEEEkeywords}
	
	\deffootnote[2em]{2em}{1em}{\textsuperscript{\thefootnotemark}\,}
	
	\section{Introduction}\label{sec:intro}
	
	Acute myocardial infarction (MI) is the most severe coronary artery disease, leading to more than a third of deaths in developed nations annually \cite{reed2017acute}. 
	It could cause irreversible damage to the myocardium by acute ischemia \cite{thygesen2018fourth}. 
	Cardiac magnetic resonance (CMR) imaging, such as late gadolinium enhancement (LGE), has been established as the clinical standard for myocardial damage evaluation \cite{guo2021cine}.
	Delineating the damaged regions of myocardial muscles \hly{could assist} MI risk stratification and treatment decision-making. 
	However, manual delineation is extremely time-consuming due to the variation of shape, location and intensity \hly{of damaged regions}. 
	Automatic segmentation approaches therefore have attracted more research attention.
	
	Conventional automatic methods, such as the signal threshold to reference mean  \cite{kolipaka2005segmentation}, full-width at half-maximum \cite{amado2004accurate}, region growing  \cite{alba2012healthy} and graph-cuts \cite{lu2012automated}, have been widely studied for myocardial infarcts  (also known as ``scars'') segmentation. 
	Recently, deep learning (DL) is widely employed to perform cardiac image analysis \cite{journal/MedIA/litjens2017,journal/MedIA/li2022}.
	Several works obtained promising accuracy via utilizing DL for myocardial pathology segmentation (MyoPS) from LGE images \cite{fahmy2020three, o2021automated, zabihollahy2019convolutional}. 
	Besides, prior knowledge was employed to further improve DL methods.
	For instance, the location information of the left \hly{ventricle (LV)} myocardium can significantly improve scar segmentation results due to the spatial relationship between scars and \hly{myocardium}. Zabihollahy \etal proposed a scar segmentation method for LGE images, their method relied on the  delineation of \hly{myocardium} contours \cite{zabihollahy2019convolutional}. 
	Note that automatically extracting \hly{myocardium} from LGE images remains a challenge, due to the heterogeneous intensity distribution of \hly{myocardium} \cite{zhuang2018multivariate}. 
	\hly{Researchers} thus developed a shape reconstruction mechanism \cite{yue2019cardiac} or \hly{model ensemble strategies \cite{zabihollahy2020fully,lin2022cascaded}} to mitigate the affect of abnormal intensity.

\begin{figure}[htbp]
		\centering
		\includegraphics[width=1\textwidth]{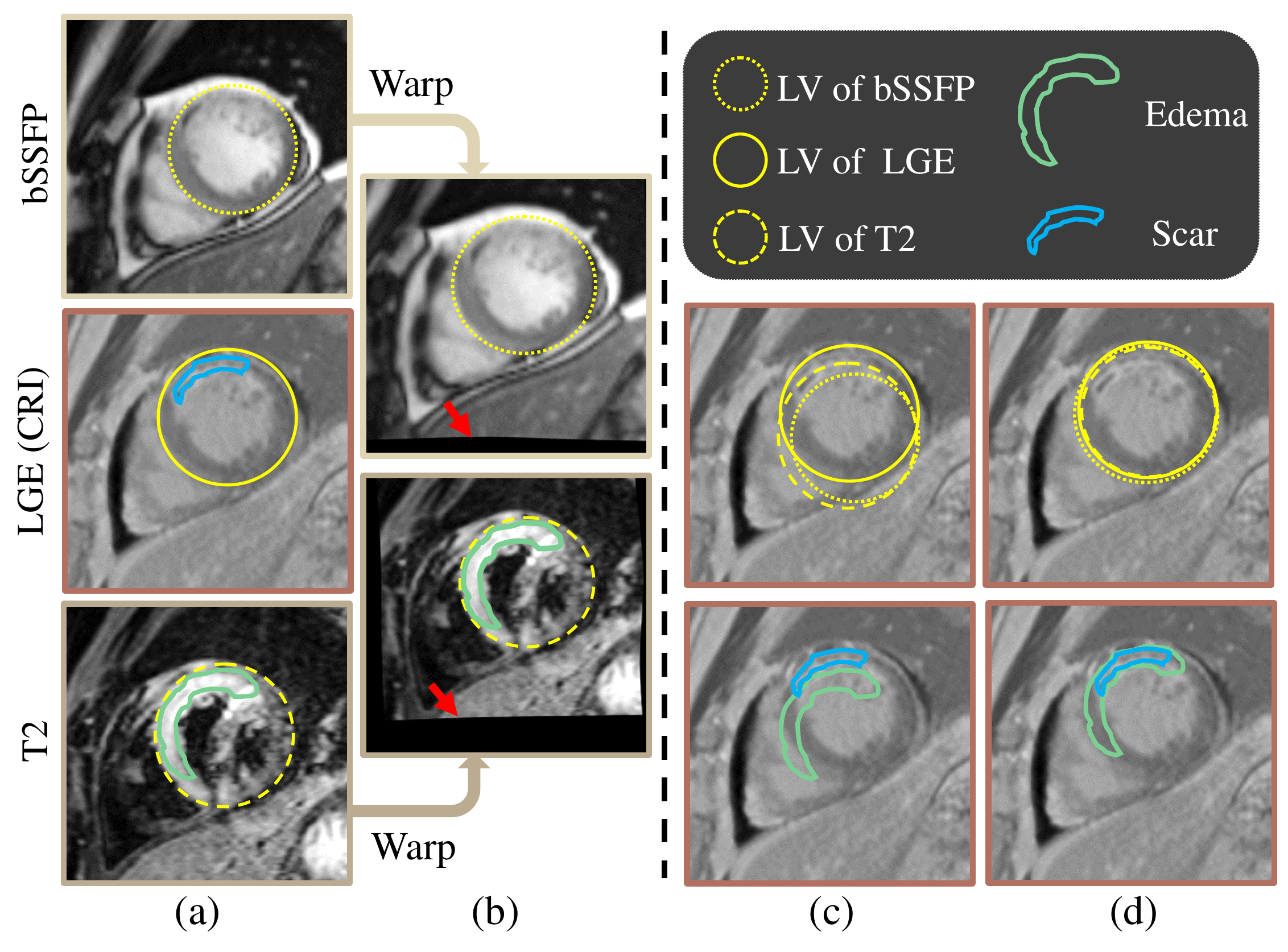}
		\caption{Visualization of unaligned and aligned multi-sequence cardiac
			magnetic resonance \hly{ (MS-CMR)} images. (a) \hly{Short-axis} views of unaligned \hly{MS-CMR} images, \ie \hly{bSSFP} (top), LGE (middle) and T2 (bottom), with contours of the left ventricle (LV) epicardium, scar and edema regions. (b) Warped \hly{bSSFP} and T2 images with contours of the LV epicardium and edema regions. Here, \hly{bSSFP} and T2 images are registered to LGE via our proposed method. \hly{One can follow red arrows to observe the differences between original and warped images.} (c) Stack of the label contours of unaligned CMR images. (d) Stack of the label contours of aligned CMR images.
			In sub-figure (c) and (d), we overlay the LV epicardium and myocardial pathology (scars and edema) contours from different CMR images on \hly{LGE}. \hly{One can that observe LV epicardium contours of original source images are initially misaligned [see (c)], and become aligned  after registering [see (d)]. CRI (common reference image) indicates the selected reference image for multi-sequence registration.}
		}
		\label{fig:intro}
	\end{figure}

	In addition to mono-sequence (such as LGE) CMR images, researchers have developed methods to process multi-sequence CMR \hly{(MS-CMR)} images.  
	Clinically, \hly{MS-CMR} images can provide various information about the heart \cite{li2022multi}. 
	\hly{For instance, balanced steady-state free precession (\hly{bSSFP})  cine  sequences present clear \hly{myocardium} boundaries [see $I_{\hly{bSSFP}}$ in \MyFig \ref{fig:intro} (a)], 
		T2-weighted (T2) CMR images provide incremental diagnostic and prognostic edema information for the scars of LGE CMR images \cite{friedrich2010myocardial} [see $I_{LGE}$ and  $I_{T2}$ in \MyFig \ref{fig:intro} (a)].
		Combining these MS-CMR images could provide relevant and complementary information  for clinical analysis, which is critical for the diagnosis and treatment management of MI diseases \cite{MyoPS2020:Online,zhuang2018multivariate,li2022myops_new_version}.}
	\hly{Moreover, utilizing MS-CMR images could also potentially promote the performance of MyoPS methods.}
	\hly{For instance, Fahmy \etal verified that adopting bSSFP  could further improve  the scar segmentation on  LGE CMR images  \cite{fahmy2021improved}.}
	To fully explore the potential advantages of using \hly{MS-CMR} images, Zhuang \etal first organized MYOPS2020 challenge\footnote{http://www.sdspeople.fudan.edu.cn/zhuangxiahai/0/myops20/}  for the design of MyoPS methods \cite{zhuang2020myocardial}. 
	

	A common limitation of current DL-based MyoPS methods is that they only can process pre-aligned \hly{MS-CMR} images \cite{wang2022awsnet,li2022taunet,fahmy2021improved}.
	Meanwhile, the only available public \hly{MS-CMR} dataset (MYOPS2020 challenge dataset) for MyoPS has  been pre-aligned prior to release.
	In clinical applications, there may exist misalignment among \hly{MS-CMR} images due to respiratory motions [see \MyFig \ref{fig:intro} (a) and (c)]. 
	For unaligned \hly{MS-CMR} images, the spatial relationship of anatomy and pathology regions protruded by different CMR images, could not be maintained consistently. 
	Therefore, integrating  information from different CMR images into a common space is desired, which could facilitate the monitoring [see \MyFig \ref{fig:intro} (a) and (d)] and  diagnosis  of MI diseases\cite{friedrich2010myocardial}.

	\hly{ This work aims to perform MyoPS for unaligned MS-CMR images from practical clinics, which segments scar, edema and healthy myocardium within one unified result for each patient. 
		We present a combined computing framework, referred to as U-MyoPS, for MyoPS \cite{zhuang2018multivariate}. U-MyoPS achieves MS-CMR image registration and myocardial segmentation and MyoPS in a unified DL-based schema. Specifically,  we propose a multi-sequence fusion (MSF) block with a TPS model to fuse unaligned MS-CMR information for myocardial segmentation.}  \hly{We also fuse unaligned pathology information based on TPS registration in U-MyoPS, facilitating pathology segmentation. Moreover,  we introduce a spatial prior gate (SPG) to  propagate myocardium information to promote the performance of pathology segmentation.  }

	
	\begin{figure*}[htb] 
		\centering
		\includegraphics[width=1\textwidth]{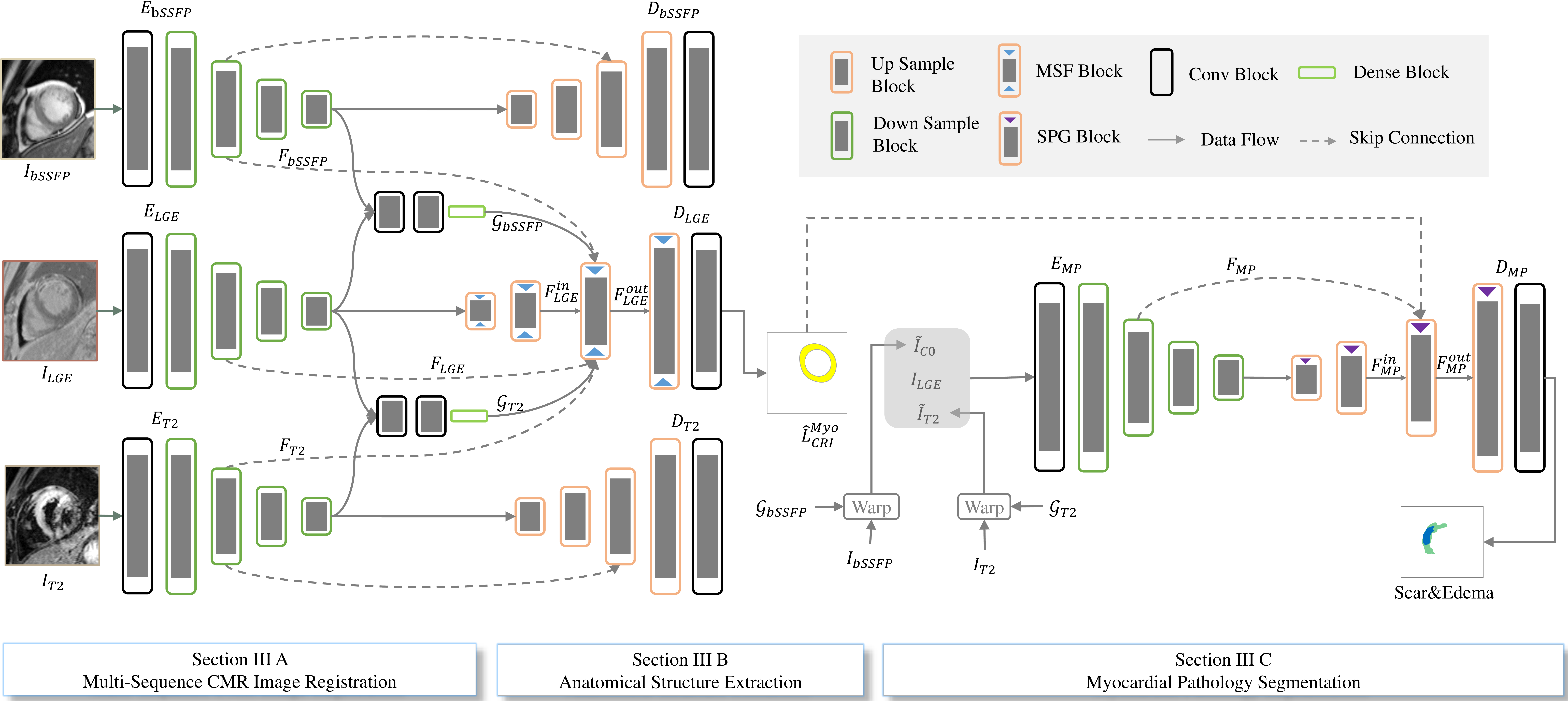}
		\caption{Pipeline of the myocardial pathology segmentation (MyoPS) framework,  \hly{referred to as U-MyoPS,} for unaligned multi-sequence \hly{cardiac
				magnetic resonance (MS-CMR)} images. 
			The network architecture of the U-MyoPS includes three encoders ($E_{\hly{bSSFP}}$, $E_{LGE}^{}$ and $E_{T2}$) and two registration heads ($R_{\hly{bSSFP}}$ and $R_{T2}$) for \hly{MS-CMR} registration,  three decoders ($D_{\hly{bSSFP}}$, $D_{LGE}^{}$ and $D_{T2}$)  for \hly{MS-CMR} anatomical structure extraction,  and a prior-aware network ($E_{MP}$ and $D_{MP}$) for MyoPS. 
			Note that we only retain representative connections between  encoders ($E_{\hly{bSSFP}}$, $E_{LGE}^{}$, $E_{T2}$ and $E_{MP}$) and decoders ($D_{\hly{bSSFP}}$, $D_{LGE}^{}$, $D_{T2}$ and $D_{MP}$), and omit the rest ones for better visualization. 
			This figure illustrates when U-MyoPS takes three  CMR images $\mathcal{I}=\{I_{\hly{bSSFP}}, I_{LGE}, I_{T2}\}$ as input, and sets $I_{LGE}$ as the common reference image (CRI) for registration and MyoPS. 
		}
		\label{fig:framework}    
	\end{figure*}
	
	\section{Method}
	
	\MyFig \ref{fig:framework} shows the pipeline of U-MyoPS, which is a combined computing method to segment myocardial pathologies from \hly{MS-CMR images}  while align them simultaneously. Initially, the input \hly{bSSFP}, T2 and LGE CMR \hly{images} are registered into a common space (Section \ref{sec:coreg}). Then, an anatomical structure  where pathologies may exist is extracted as the spatial prior (Section \ref{sec:prior}). Finally, U-MyoPS segments myocardial pathology regions inside the spatial prior (Section \ref{sec:seg}).

	\subsection{\hly{MS-CMR} Image Registration}
	\label{sec:coreg}
	
	We align \hly{MS-CMR} images into a common space via neural networks. 
	Let $\mathcal{I}=\{I_{\hly{bSSFP}},I_{LGE},I_{T2}\}$ denotes a set of 2D  \hly{MS-CMR} images extracted from the CMR data of the same subject (see Section \ref{sec:dataset} for extraction process). 
	The size of each extracted image is $H \times W$.  
	To align these images, we can set one of them as the common reference image (CRI), and register the rest images to it.
	For convenience, we set $I_{LGE}$ as the CRI in this section.
	In U-MyoPS, we introduce three encoders ($E_{\hly{bSSFP}}$, $E_{LGE}^{}$ and $E_{T2}$) to capture underlying structural information from $\mathcal{I}$, and two registration heads ($R_{\hly{bSSFP}}$ and $R_{T2}$) to estimate TPS transformations for registration.

	A TPS transformation is parameterized via a grid of control points \cite{rocco2017convolutional}. \hly{Briefly}, we set an imaginary grid of control points on $I_{\hly{bSSFP}}$ and $I_{T2}$, and warp the images according to the displacements of control points. In U-MyoPS, the grid is defined with $m\times m$ equally spaced control points.
	$R_{\hly{bSSFP}}$ and $R_{T2}$ predict the displacements of each control point:  
	\begin{equation}
		\begin{split}
			\mathcal{G}_{\hly{bSSFP}}=\{(\delta x^{\text{\tiny{\hly{bSSFP}}}}_k, \delta y^\text{\tiny{\hly{bSSFP}}}_k)| k=1...m\times m\},\\
			\mathcal{G}_{T2}=\{(\delta x^\text{\tiny{T2}}_k, \delta y^\text{\tiny{T2}}_k)| k=1...m\times m\},\\
		\end{split}
		\label{eq:tps_param}
	\end{equation}
	where $(\delta x^\text{\tiny{\hly{bSSFP}}}_k, \delta y^\text{\tiny{\hly{bSSFP}}}_k)$ and $(\delta x^\text{\tiny{T2}}_k, \delta y^\text{\tiny{T2}}_k)$ are the displacements of control point $(x_k, y_k)$.  
	$I_{\hly{bSSFP}}$ and  $I_{T2}$ can be warped to  $I_{LGE}$ \hly{by the predicted displacements as follows}:
	\begin{equation}
		\begin{split}
			\tilde{I}_{\hly{bSSFP}} & = \mathcal{T}_{\mathcal{G}_{\hly{bSSFP}}^{}}(I_{\hly{bSSFP}}),    \\
			\tilde{I}_{T2} & = \mathcal{T}_{\mathcal{G}_{T2}^{}}(I_{T2}),  
		\end{split}
		\label{eq:transform}    
	\end{equation}
	where $\tilde{I}_{\hly{bSSFP}}$ and $\tilde{I}_{T2}$ are the warped images, $\mathcal{T}_{\mathcal{G}_{\hly{bSSFP}}^{}}$ and $\mathcal{T}_{\mathcal{G}_{T2}^{}}$ denote the \hly{mapping function}, 
	\hly{whose parameters were determined by ${\mathcal{G}_{\hly{bSSFP}}^{}}$ and ${\mathcal{G}_{T2}^{}}$ via existing \hly{DL-based closed-form solution}  \cite{jaderberg2015spatial,bookstein1989principal}, respectively.}
	After registration, we can compensate the misalignment of $\mathcal{I}$, and obtain aligned \hly{MS-CMR} images $\mathcal{I}_{}^{\prime}=\{\tilde{I}_{\hly{bSSFP}}, I_{LGE}, \tilde{I}_{T2}\}$ in the common space. Consequently, the spatial relationship of anatomy and pathology regions protruded by different CMR images could be consistent in $\mathcal{I}_{}^{\prime}$, promoting the subsequent anatomical structure extraction and pathology segmentation tasks. 
	
	\textbf{Registration Loss:}
	To optimize the parameters of the encoders and registration heads, we introduce a multi-sequence registration loss function as follows:
	\begin{equation}
		\begin{split}
			\mathcal{L}oss_{Reg}= -\sum \limits_{ j \in \{\hly{bSSFP},T2\}}\mathcal{D}ice(\mathcal{T}_{\mathcal{G}_{j}^{}}({L}^{ana}_{j}), 
			{L}^{ana}_{LGE}),
		\end{split}
		\label{eq:reg_loss}
	\end{equation}
	where $\mathcal{D}ice(a,b)$ calculates the Dice score between label $a$ and $b$ \cite{hu2018weakly},  $L_{j}^{ana}$ and $L_{LGE}^{ana}$  denote the corresponding anatomical structure labels, \hly{\ie myocardium, LV and right ventricle (RV),} of \hly{MS-CMR} images.  
	With the supervision of $\mathcal{L}oss_{Reg}$, registration headers could learn to align \hly{MS-CMR} images.

	\subsection{Anatomical Structure Extraction }\label{sec:prior}
	
	We introduce a decoder $D_{LGE}^{}$ to extract the anatomical structure, \ie \hly{myocardium,} of CRIs.
	Generally, $D_{LGE}$ could utilize feature maps from $E_{LGE}$ to predict \hly{myocardium} regions. 
	Note that the feature maps from $E_{\hly{bSSFP}}$ and $E_{T2}$  also provide \hly{myocardium} structure information, we further propose MSF to fuse these information into $D_{LGE}$ to improve \hly{myocardium} extraction. 
	
	\begin{figure}[htb] 
		\centering
		\includegraphics[width=0.92\textwidth]{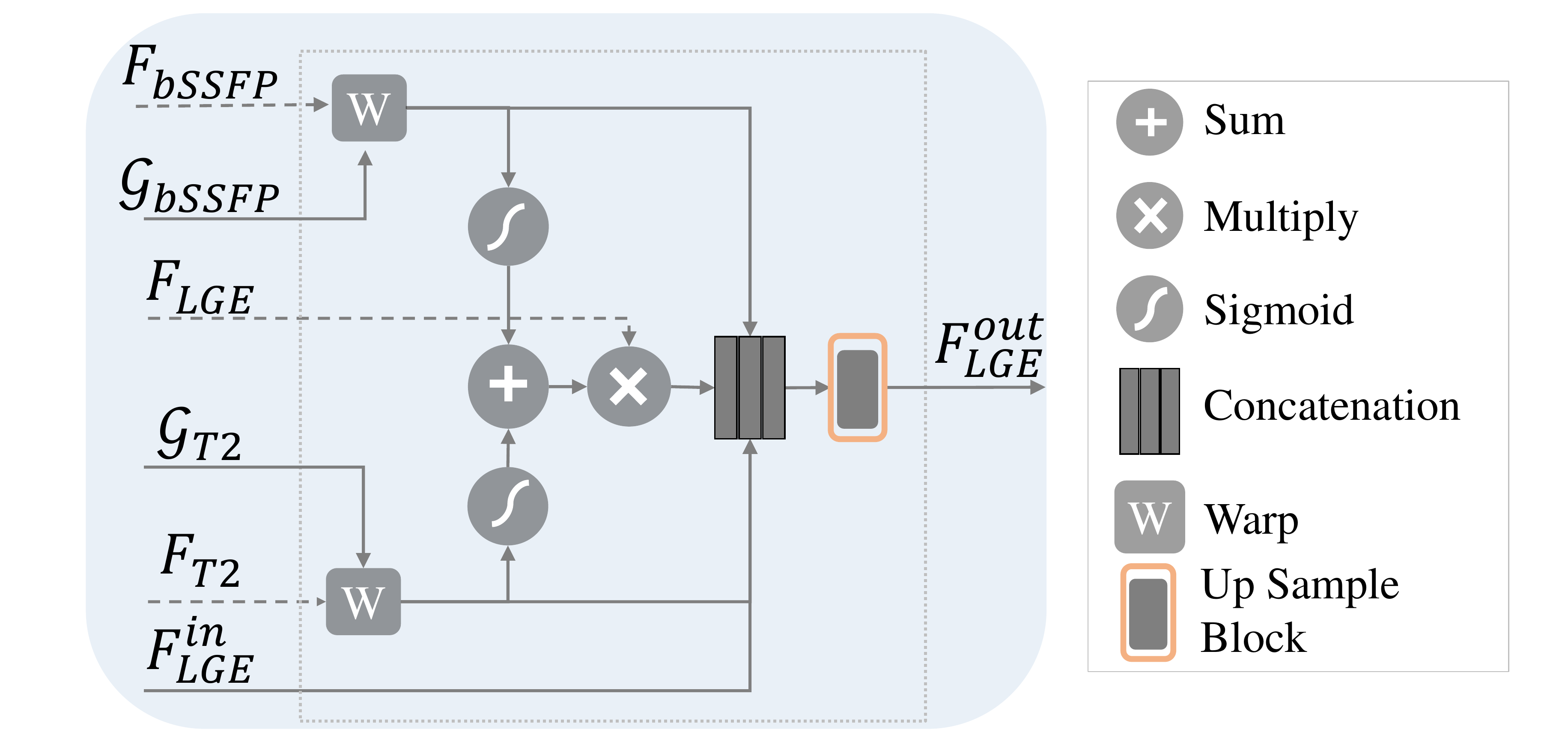}
		\caption{Schematic illustration of proposed multi-sequence fusion (MSF) \hly{block} in $D_{LGE}$. }
		\label{fig:MSF}
	\end{figure}
	
	\MyFig \ref{fig:MSF} visualizes MSF \hly{block}. 
	The main idea of MSF lies in compensating spatial misalignment before feature map fusion. 
	Let ${F}_{LGE}^{in}$ be the previous output in $D_{LGE}^{}$, while
	${F}_{\hly{bSSFP}}^{}$, ${F}_{T2}^{}$ and ${F}_{LGE}^{}$ be the feature maps at $i$-th level of $E_{\hly{bSSFP}}$, $E_{T2}$ and $E_{LGE}$, respectively. 
	\hly{Since $I_{bSSFP}$, $I_{T2}$ and $I_{LGE}$ were unaligned images, the corresponding feature maps $F_{bSSFP}$, $F_{T2}$ and $I_{LGE}$ remained unaligned until input into MSF module. Thus, ${F}_{\hly{bSSFP}}^{}$ and ${F}_{T2}^{}$ could not be directly fused with ${F}_{LGE}^{}$.}
	Here, we employ TPS transformations to re-calibrate  the feature maps for fusion. Note that the spatial sizes of feature maps vary across different levels of encoders due to down-sample operations (\ie max pooling). To transform the feature maps, we first \hly{adjust} TPS parameters as follows:
	\begin{equation}
		\begin{split}
			\mathcal{G}^{h\times w}_{\hly{bSSFP}}=\{(\delta x^\text{\tiny{\hly{bSSFP}}}_k\hly{\times}h/H, 
			\delta y^\text{\tiny{\hly{bSSFP}}}_k\hly{\times}w/W) | k=1...m \times m\},\\
			\mathcal{G}^{h\times w}_{T2}=\{(\delta x^\text{\tiny{T2}}_k\hly{\times}h/H, 
			\delta y^\text{\tiny{T2}}_k\hly{\times}w/W) | k=1...m \times m\},
		\end{split}
		\label{eq:feat_rescale}
	\end{equation}
	where $h\times w$ and $H \times W$ denote the spatial size of feature maps and original CMR images, respectively. Then, ${F}_{\hly{bSSFP}}^{}$ and ${F}_{T2}^{}$ can be warped via $\mathcal{G}^{h\times w}_{\hly{bSSFP}}$ and $\mathcal{G}^{h\times w}_{T2}$, respectively (see \eqref{eq:transform}). Finally, we concatenate the warped ${F}_{\hly{bSSFP}}^{}$ and ${F}_{T2}^{}$ together with ${F}_{LGE}^{}$ and $F_{LGE}^{in}$. The output of MSF ${F}^{out}_{LGE}$ contains \hly{myocardium} information from \hly{MS-CMR} images, which could improve the extraction of \hly{myocardium} structures.

	\textbf{\hly{Myocardium} Extraction Loss:}
	The extracted \hly{myocardium} structure should be consistent with the \hly{myocardium} contour of CRI images. To optimize the parameter of $D_{LGE}$, we introduce a \hly{myocardium} extraction loss  as follows: 
	\begin{equation}
		\mathcal{L}oss_{Myo}=-\mathcal{D}ice(\hat{L}^{Myo}_{LGE},{L}^{Myo}_{LGE}),
	\end{equation}
	where $\hat{L}^{Myo}_{LGE}$ and ${L}^{Myo}_{LGE}$ are the extracted \hly{myocardium} and gold standard \hly{myocardium} label of $I_{LGE}$ (CRI). Besides, we introduce two additional decoders $D_{\hly{bSSFP}}$ and $D_{T2}$  to capture myocardial information from $I_{\hly{bSSFP}}$ and $I_{T2}$, respectively. $D_{\hly{bSSFP}}$ and $D_{T2}$ are trained via \hly{myocardium} constraint loss as follows: 
	\begin{equation}
		\mathcal{L}oss_{Cons}=-\sum_{j \in \{\hly{bSSFP},T2\}}\mathcal{D}ice(\hat{L}^{Myo}_{j},{L}^{Myo}_{j}),
		\label{eq:cons}
	\end{equation}
	where $\hat{L}^{Myo}_{j}$ and ${L}^{Myo}_{j}$ are the predicted \hly{myocardium} and gold standard labels of CMR images, respectively. With the constraint loss, $E_{\hly{bSSFP}}$ and $E_{T2}$ could focus on capturing myocardial features from $I_{\hly{bSSFP}}$ and $I_{T2}$, facilitating the \hly{myocardium} extraction of $D_{LGE}^{}$.
	
	Furthermore, we simultaneously train anatomical structure extraction with multi-sequence registration tasks via a hybrid loss function:
	\begin{equation}
		\mathcal{L}oss_{Hyb}=\mathcal{L}oss_{Reg}+ \lambda ( \mathcal{L}oss_{Cons}+\mathcal{L}oss_{Myo}),
		\label{eq:hybrid}
	\end{equation}
	where $\lambda$ is the balance coefficient for multi-sequence registration,  \hly{myocardium} extraction and \hly{myocardium} constraint losses.
	\hly{Here, U-MyoPS performs MS-CMR image registration and myocardium  segmentation simultaneously and collaboratively, and optimizes network parameters by merging loss functions, which is similar to existing joint optimization methods \cite{luo2020mvmm,upendra2021joint,qin2018joint}. Nevertheless, U-MyoPS further introduces a novel network module, \ie MSF, to promote joint optimization tasks.}

	\subsection{Myocardial Pathology Segmentation}\label{sec:seg}

	With extracted \hly{myocardium} ($\hat{L}^{Myo}_{LGE}$), we construct a prior-aware sub-network to segment edema and scars from $\mathcal{I}_{}^{\prime}$. The backbone of the network is a U-Shape model \cite{ronneberger2015u}, including \hly{an} encoder ($E_{MP}$) and a decoder ($D_{MP}$). 
	Notably, $\hat{L}^{Myo}_{LGE}$  provides spatial prior for myocardial pathologies. We propose SPG in $D_{MP}$ to incorporate $\hat{L}^{Myo}_{LGE}$ for pathology segmentation.
	
	\begin{figure}[htb] 
		\centering
		\includegraphics[width=0.92\textwidth]{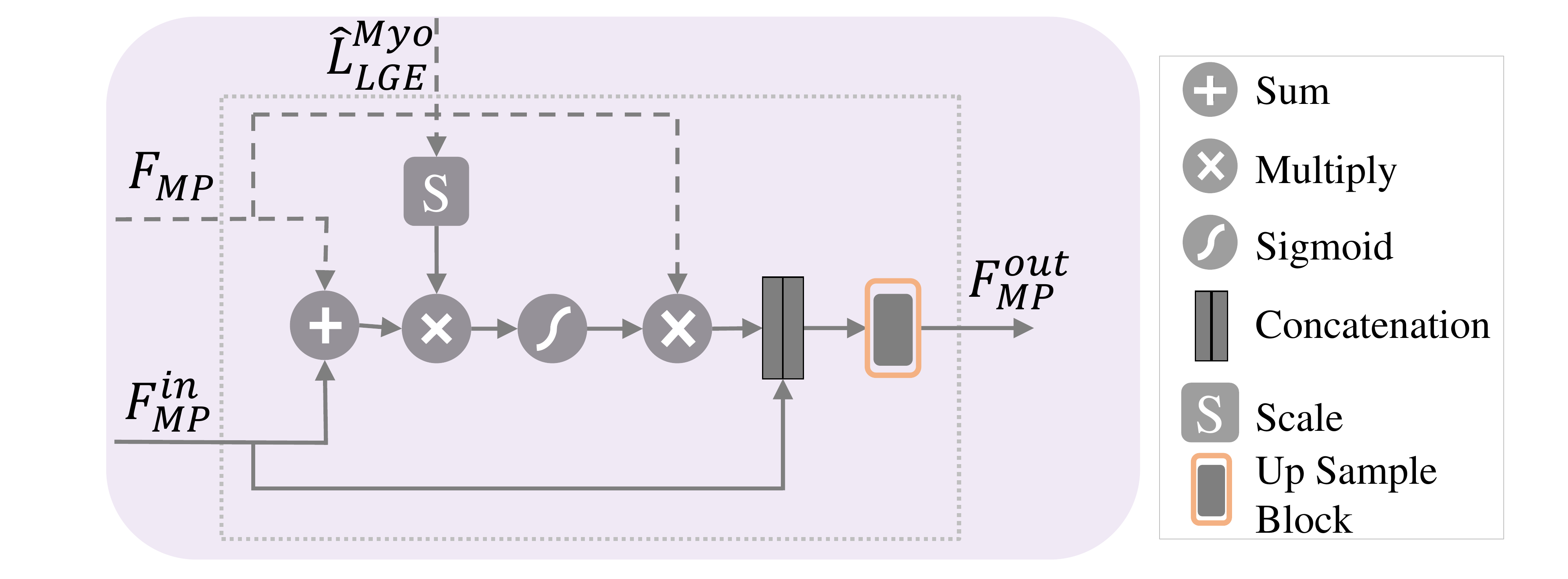}
		\caption{Schematic illustration of the proposed spatial prior gate (SPG) \hly{block} in $D_{MP}$}
		\label{fig:SPG}
	\end{figure}
	\MyFig \ref{fig:SPG} shows SPG \hly{block}, \hly{which aims to propagate myocardium information to promote MyoPS.}
	Let  ${F}_{MP}^{in}$ be previous output in $D_{MP}$, and ${F}_{MP}^{}$ denotes a feature map  from $i$-th level of $E_{MP}$. 
	\hly{SPG could highlight informative regions of ${F}_{MP}^{}$, and aggregate the information of recalibrated ${F}_{MP}^{}$ and ${F}_{MP}^{in}$ for pathology prediction. }
	Specifically, we first add $F_{MP}$ with ${F}_{MP}^{in}$, and highlight their informative region by multiplying with $\hat{L}^{Myo}_{LGE}$. Note that the spatial sizes of ${F}_{MP}^{}$ and ${F}_{MP}^{in}$  vary in different levels, and we thus employ a scale operation to adjust $\hat{L}^{Myo}_{LGE}$ to different sizes of ${F}_{MP}^{}$ and ${F}_{MP}^{in}$. After then, sigmoid operations are applied to obtain an attention map. Finally, we multiply ${F}_{MP}^{}$ with the  attention map, and concatenate  the multiplication result with  ${F}_{MP}^{in}$ to obtain the output of SPG ${F}_{MP}^{out}$.

	\textbf{Myocardial Pathology Segmentation Loss:} We train the prior-aware sub-network by minimizing the difference between predicted and gold standard myocardial pathology labels:
	
	\begin{equation}
		\begin{split}
			\mathcal{L}oss_{MP}=-\mathcal{D}ice(\hat{{L}}_{}^{\prime},{L}_{}^{\prime})+\mathcal{C}E(\hat{{L}}_{}^{\prime},{L}_{}^{\prime}),
		\end{split}
		\label{eq:loss:seg}
	\end{equation}
	where $\hat{{L}}_{}^{\prime}$ and  ${L}_{}^{\prime}$  are the predicted and gold standard pathology labels of $\mathcal{I}_{}^{\prime}$, respectively; $\mathcal{C}E(a,b)$ calculates the cross-entropy between $a$ and $b$.
	Note that scar and edema labels were only originally delineated in unaligned LGE and T2 images, respectively  \hly{[see the lower part of \MyFig \ref{fig:intro} (c)]}. 
	We align the original scar and edema labels into a common space to obtain the gold standard pathology label of $\mathcal{I}_{}^{\prime}$. 
	Namely, ${L}_{}^{\prime}$ is a combination of scar and edema labels as shown in  the lower part of \MyFig \ref{fig:intro} (d). \hly{For the overlap part of the  scar and edema labels, we denote them as scar regions.  For the non-overlapped parts, we retained to use the original label.}
	
	Finally, the overall parameters of U-MyoPS, 
	\ie $E_{\hly{bSSFP}}$, $E_{LGE}$, $E_{T2}$, $D_{\hly{bSSFP}}$, $D_{LGE}$, $D_{T2}$, $R_{\hly{bSSFP}}$, $R_{T2}$, $E_{MP}$ and $D_{MP}$, can be optimized by minimizing $\mathcal{L}oss_{Hyb}$ and
	$\mathcal{L}oss_{MP}$.

	\section{Experiments}
	
	\subsection{Datasets and Experiment Settings}\label{sec:dataset}
	
	We evaluated the proposed framework, \ie U-MyoPS, on a private unaligned \hly{MS-CMR} dataset (pMM-CMR dataset) and a public pre-aligned  \hly{MS-CMR} dataset (MYOPS2020 challenge dataset).
	\begin{itemize}
		\item pMM-CMR dataset: This dataset includes 50  unaligned \hly{MS-CMR} (\ie bSSFP, LGE and T2)  images.  
		\hly{The typical  imaging parameters of  bSSFP  were  as follows: field of view (FOV) = 283 $\sim$ 453 $\times$ 283$\sim$453  mm$^2$, pixel size=  1.47$\sim$2.36$\times$ 1.47$\sim$2.36  mm$^2$ , number of slices= 8$\sim$10, slice thickness= 10 mm. 
			The  imaging parameters of LGE were:  FOV = 283$\sim$453 $\times$ 283$\sim$453  mm$^2$, pixel size =  1.33$\sim$1.86 $\times$ 1.33$\sim$1.86  mm$^2$,  number of slices= 4$\sim$11, and slice thickness = 10 mm. 	
			The imaging parameters of T2 were as follows: FOV = 283$\sim$453 $\times$ 283$\sim$453   mm$^2$, pixel size =  1.11$\sim$1.77 $\times$  1.11$\sim$1.77  mm$^2$, number of slice= 2$\sim$10, and slices thickness =  10 mm. }
		
		\hly{We employed the end-diastolic cardiac phase of bSSFP cine data, which is consistent to the cardiac phase of LGE and T2.} 
		\hly{The ground truth contours were annotated by one cardiologist. The delineation was performed by slice-wise using  ITK-SNAP software \cite{yushkevich2006user}. For each slice, the cardiologist acquired labels as follows: (1) The cardiologist delineated the structure labels, \ie LV, RV and myocardium. (2) The cardiologist manually adjusted gray-scale to identify all  highlighted areas within the myocardium as scars in LGE images \cite{fahmy2021improved}. If necessary, the cardiologist manual included scars that were  not identified by the intensity threshold (such as microvascular obstruction) \cite{fahmy2021improved}. (3) The cardiologist manually adjusted gray-scale to identify all highlighted areas within the myocardium as edema in T2 images. If necessary, the cardiologist manual excluded areas representing noise or artefacts (such as suppressed blood signal) \cite{eitel2011t2}.}
		In experiments, we randomly selected 20 unaligned \hly{MS-CMR} images for network training and validation,  while leaved the rest 30 for testing.

	\end{itemize}
	\begin{itemize}
		\item MYOPS2020 challenge dataset: This dataset contains 45 (25 labeled and 20 unlabeled)  \hly{MS-CMR}  (\ie bSSFP, LGE and T2)  images  \cite{li2022myops_new_version}. All released images have been aligned into a common space via MvMM \cite{zhuang2018multivariate}.	Similar to pMM-CMR dataset, the labeled CMR images were manually delineated with scar, edema, \hly{myocardium,} LV and RV labels. In experiments, we used the labeled images for network training and validation, and \hly{tested our method on unlabeled images with MYOPS2020 evaluation kit\footnote{\url{http://www.sdspeople.fudan.edu.cn/zhuangxiahai/0/myops20/test_image_evaluation_kit.zip}}.} 
	\end{itemize}
	
	\begin{table}
		
		\centering
		\caption{\hly{The demographic information of MYOPS2020 and pMM-CMR datasets.  We randomly split the images of pMM-CMR to generate training and test set. No: number of images. M/ F: male/ female. MI: myocardial infarction.}}
		\resizebox{\textwidth}{!}{
			\begin{tabular}{l|ccccc}
				\hline
				\hly{Dataset} & \hly{No} & \hly{Age} & \hly{M/ F} & \hly{Training/ Test} & \hly{Diagnoisis} \\ \hline
				\hly{MYOPS2020}   & \hly{45}  & \hly{56.2 $\pm$ 7.92}    &  \hly{45M/ 0F}   & \hly{20/ 25} & \hly{acute MI}           \\
				\hly{pMM-CMR} & \hly{50}  &   \hly{55.6 $\pm$ 8.69}    &   \hly{48M/ 2F}     & \hly{20/ 30} & \hly{acute MI}    \\ \hline
			\end{tabular}
		}
		\label{tab:demograph}
	\end{table}
	
\begin{figure}[h] 
		\centering
		\includegraphics[width=1\textwidth]{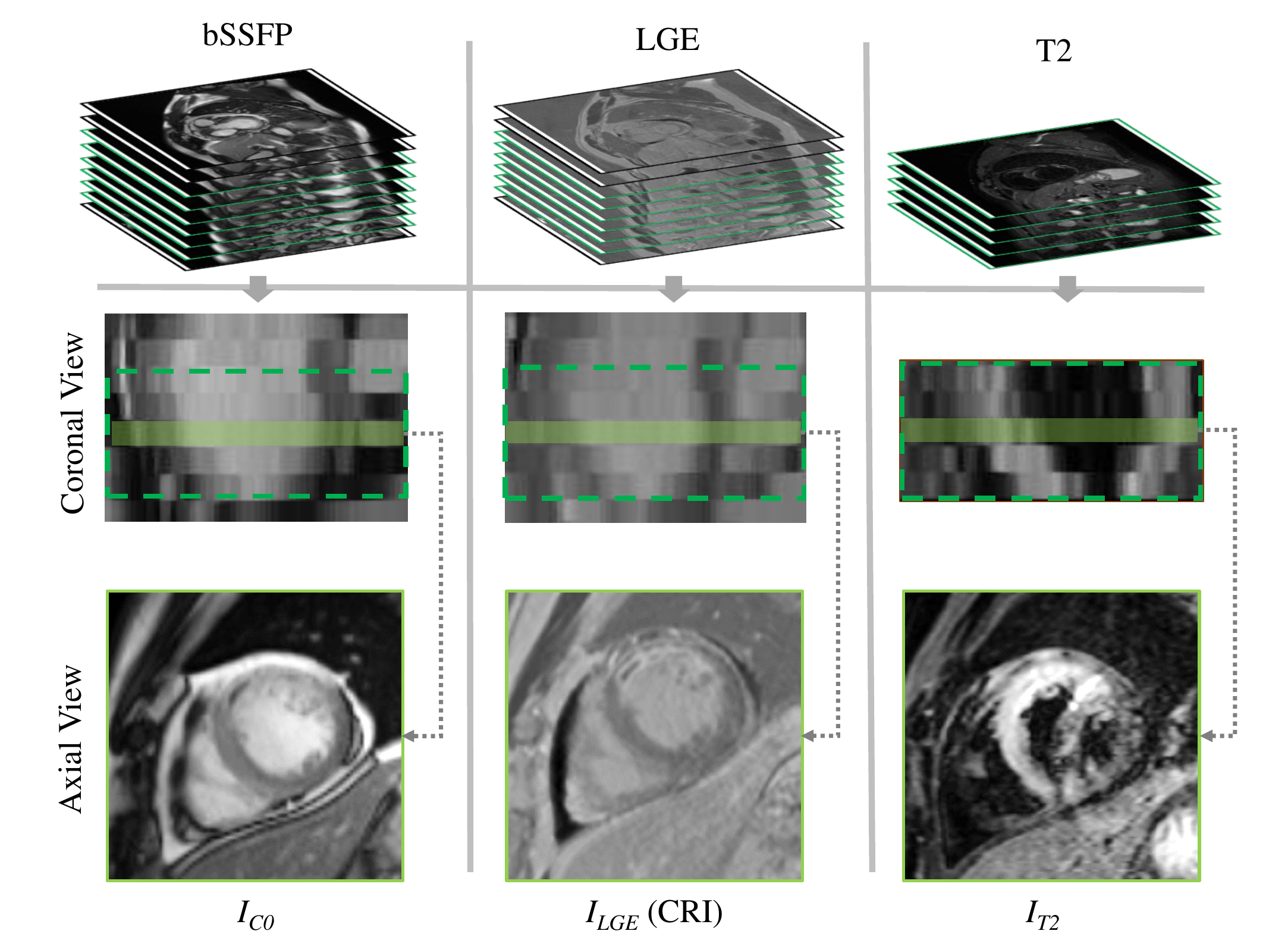}
		\caption{Illustration of multi-sequence cardiac magnetic resonance  image extraction. Green \hly{dashed} rectangles mark the physical regions that are simultaneously imaged by \hly{bSSFP}, T2 and LGE. \hly{ CRI (common reference image) indicates the selected reference image for multi-sequence registration.}}
		\label{fig:slice:extrac}    
	\end{figure}

	\hly{pMM-CMR and MYOPS2020 dataset are  independent datasets from different clinical centers.} 
	\hly{Table \ref{tab:demograph} lists the demographic information of two datasets.}
	\hly{The original CMR images of the two datasets are breath-hold multi-slice 2D images, acquired along the cardiac short-views.}
	We extracted slices from the original images of the same subject for network training and testing.
	\MyFig \ref{fig:slice:extrac} illustrates the extraction procedure.  Given \hly{bSSFP}, T2 and LGE images of the same subjects (see the top row of \MyFig \ref{fig:slice:extrac}), we first pre-aligned these CMR images via rigid transformation. Note that the field-of-views may vary across different sequences. For example, \hly{bSSFP}  usually visualizes the whole LV, while T2  may only cover the sub-region of LV.
	We then employed the slices from the physical regions that \hly{were} simultaneously imaged by \hly{bSSFP}, T2 and LGE (see the middle row of \MyFig \ref{fig:slice:extrac}). 
	After then, by regarding LGE slice ($I_{LGE}$) as the common reference image, we extracted the corresponding \hly{bSSFP} ($I_{\hly{bSSFP}}$) and T2 ($I_{T2}$) slices which have the closest physical distance to it (see the bottom row of \MyFig \ref{fig:slice:extrac}). 
	\hly{ Note that to calculate the physical distance between CMR slices, we directly employed the transformation parameters of  the images from the header of NIFIT data\cite{li2021right}.}
	Finally, all slices were cropped by centering at the heart regions. 
	Each extracted slice was re-sampled with the same spatial resolution, and normalized \hly{via} Z-score. 
	\textit{We primarily evaluated on the pMM-CMR dataset (see Sec. \ref{exp:results:pMM-CMR:comparasion and ablation study}, \ref{sec:seg:scar:eval},\ref{exp:results:pMM-CMR:num of sequence}, \ref{exp:results:pMM-CMR:correlation} and \ref{sec:reg:res}), while the MYOPS2020 challenge dataset was only used for a comparison with the state-of-the-art methods on a public dataset (see Sec. \ref{exp:results:MYOPS2020}). }

	\subsection{Network Training}\label{sec:train}

	U-MyoPS was implemented by Pytorch and trained with Adam optimizer\footnote{Source codes will be released publicly at \url{https://github.com/NanYoMy/myops} once the manuscript is accepted.}. For \hly{MS-CMR} image registration,  TPS grids were \hly{initially} set with $4 \times 4$ \hly{equally-spaced} control points:
	\hly{ 
		\begin{equation}
			\mathcal{G}_C=\{(x_k,y_k)| k=1,\dots, 16\},
		\end{equation}
		where $x_k,y_k \in \{p|p=256 \times (-0.98+0.65 \times n) , n= 0,1,2,3\}$. Note that we set the center of images to (0,0), and resized images to 256$\times$256 when performing TPS transformation.} 
	To train U-MyoPS, unaligned \hly{MS-CMR images} were fed into the network. We first jointly trained the multi-sequence registration and anatomical structure extraction with the hybrid loss [see \eqref{eq:hybrid}] by setting $\lambda$  to 0.1. After converging, we \hly{froze} the parameters for multi-sequence registration and anatomical structure extraction, and then optimized the parameters of prior-aware sub-network by minimizing pathology segmentation loss [see \eqref{eq:loss:seg}].

	\subsection{Gold Standard and Evaluation Metrics}\label{sec:gd}
	
	We evaluated the accuracies of MyoPS in the common space.
	Note that in the final MyoPS visualization, edema labels were partially overlapped by scar labels in common spaces. 
	We refer to the union of scar and edema labels as edema in the following \hly{evaluation} and visualization \cite{li2022myops_new_version}. 
	Here, Dice and Hausdorff distance (HD) and statistical measurements [\ie Sensitivity (Sen) and Precision (Pre)] were employed for MyoPS evaluation.
	\hly{Meanwhile,  scar size, scar transmurality and edema size were included for clinical quantification.}
	\hly{Besides, Dice and HD between the warped source and target labels were adopted for multi-sequence registration evaluation. }
	

	\subsection{Results of MyoPS on pMM-CMR Dataset} \label{exp:results:pMM-CMR:comparasion and ablation study}
	\subsubsection{Comparison with \hly{Semantic Segmentation} Methods}\label{sec:compare_existing_method}
	
	We first compared U-MyoPS with  \hly{existing DL-based semantic} segmentation methods. 
	Here, we implemented  \hly{five semantic segmentation based on their official codes}\footnote{PSN:~{https://github.com/ShawnBIT/UNet-family}; \\
		nn-Unet:~{https://github.com/MIC-DKFZ/nnUNet}; \\
		MvMM:~{http://www.sdspeople.fudan.edu.cn/zhuangxiahai/0/zxhproj/};\\
		AWSnet:~{https://github.com/soleilssss/AWSnet/tree/master}.
	} as well as our method: 
	\begin{itemize}
		\item \hly{nn-Unet$^\text{Una}$: nn-Unet \cite{isensee2021nnu} which took unaligned \hly{MS-CMR images}  as inputs, and predicted scarring and edema regions. Note that the predicted scarring and edema regions were unaligned due to the misalignment among input images.} 
		
		\item PSN$_{\text{LGE}}$: U-shape \cite{ronneberger2015u} pathology segmentation network which was trained with $I_{LGE}$ as well as corresponding scar labels.  
		
		\item PSN$_{\text{T2}}$: U-shape \cite{ronneberger2015u} pathology segmentation network  which was trained with $I_{T2}$ as well as corresponding edema labels.

		\item MvMM+nn-Unet: nn-Unet \cite{isensee2021nnu} which took aligned \hly{MS-CMR images}  as inputs, and predicted scarring and edema regions. Here, the \hly{MS-CMR} images were aligned by MvMM \cite{zhuang2018multivariate}, which was also the standard tool for \hly{aligning} \hly{MS-CMR} images of MYOPS2020 challenge dataset \cite{li2022myops_new_version}. 
		
		
		\item \hly{MvMM+AWSnet: AWSnet \cite{wang2022awsnet}  which took aligned \hly{MS-CMR images}  as inputs, and predicted scarring and edema regions. Here, \hly{MS-CMR} images were also aligned by MvMM \cite{zhuang2018multivariate}. AWSnet  is a promising coarse-to-fine  MyoPS method. It first obtained LV regions from bSSFP images. Then, it segmented scars and edema  based on  the LV regions.}
		
		\item U-MyoPS$_{\text{\hly{bLT}}}$: U-MyoPS which utilizes \hly{unaligned MS-CMR images}  for scar and edema segmentation.

	\end{itemize}

	\begin{table*}[htb] 
		\resizebox{1\textwidth}{!}{ 
			\begin{tabular}{lccc|cccc|cccc}
				\hline
				\multirow{2}{*}{Method} & \multicolumn{3}{c}{Sequences} & \multicolumn{4}{|c|}{Scar }& \multicolumn{4}{c}{Edema }\\

				& \hly{bSSFP}  & LGE  & T2  & Dice (\%) $\uparrow$  & Sen  (\%) $\uparrow$ & Pre (\%) $\uparrow$ & HD (mm)   $\downarrow$ & Dice (\%) $\uparrow$  & Sen  (\%) $\uparrow$ & Pre (\%) $\uparrow$ & HD (mm)   $\downarrow$ \\
				\hline 
				\hly{nn-Unet$^{\text{Una}}$}  & \checkmark & \checkmark & \checkmark & \hly{44.16 (17.47)} & \hly{46.68 (20.30)}  & \hly{44.05 (18.11)} & \hly{37.88 (21.84)}  & \hly{66.48 (14.74)}  & \hly{70.97 (13.26)} &\hly{ 64.81 (18.47)}  & \hly{38.28 (23.40)}  \\
				
				PSN$_{\text{LGE}}$ & $\times$ &  \checkmark  & $\times$ &55.99 (17.52)  & {61.66 (20.21)} & 53.25 (18.80) & 43.01 (23.08) & \multicolumn{4}{|c}{N/A }\\

				PSN$_{\text{T2}}$ & $\times$ &  $\times$ & \checkmark & \multicolumn{4}{c|}{N/A} & 67.82 (19.01)   & {79.42 (12.86)} & 61.97 (22.94)  &31.07 (23.79) \\

				MvMM+nn-Unet  & \checkmark & \checkmark & \checkmark & 57.18 (11.18) & 62.29 (12.59) & 54.23 (13.31) & 36.93 (17.50) & 69.87 (14.13) & {78.42 (12.66)} & 65.25 (17.96) & {37.79 (22.29)} \\
				
				\hly{MvMM+AWSnet}  & \checkmark & \checkmark & \checkmark & \hly{62.38 (14.47) } & \hly{ 68.02 (18.28) } & \hly{ 59.20 (13.52) } & \hly{ 36.56 (15.53) } & \hly{ 74.23 (12.87) } & \hly{ \textbf{82.13 (8.875)}} & \hly{70.03 (17.79)} & \hly{30.12 (22.97)}  \\
				
				U-MyoPS$_{\text{\hly{bLT}}}$ & \checkmark & \checkmark & \checkmark &   \textbf{64.92 (9.816)}   & \textbf{68.30 (12.56) } & {63.34 (11.71)} & \textbf{29.16 (16.65)} & \textbf{76.01 (9.784)} & {80.49 (8.942)} & \textbf{73.53 (14.05)} & \textbf{27.89 (18.45)}\\

				\hline
				\hline
				U-MyoPS$^{\text{w/o MSF}}_{\text{\hly{bLT}}}$ & \checkmark & \checkmark & \checkmark &  64.58 (9.762)  & 64.76 (12.08) & \textbf{66.12 (11.77)} & 33.61 (18.21) & 75.15 (11.21) & 79.78 (10.56) & 72.56 (14.88) & {32.65 (21.38)} \\
				
				U-MyoPS$^{\text{w/o SPG}}_{\text{\hly{bLT}}}$ & \checkmark & \checkmark & \checkmark &   60.61 (10.11)  & 65.03 (14.21) & 58.71 (12.36) & 34.17 (19.32) & 71.69 (13.56) & 75.06 (11.97) & 70.94 (18.38) & {32.98 (19.61)} \\

				\hline
				
			\end{tabular}
			
		}{
			\caption{Performance of different MyoPS methods. Note that PSN$_\text{LGE}$ and PSN$_\text{T2}$ could only predict scar and edema of LGE and T2 images, respectively. Subscript ``$_\text{\hly{bLT}}$'': \hly{bSSFP}, LGE and T2 images.  The best results are labeled in \textbf{bold}. 
			}
			\label{tab:rj:U-MyoPS}
		}
		
	\end{table*}	
	
	\begin{figure*}[htb] 
		\centering
		\includegraphics[width=1\textwidth]{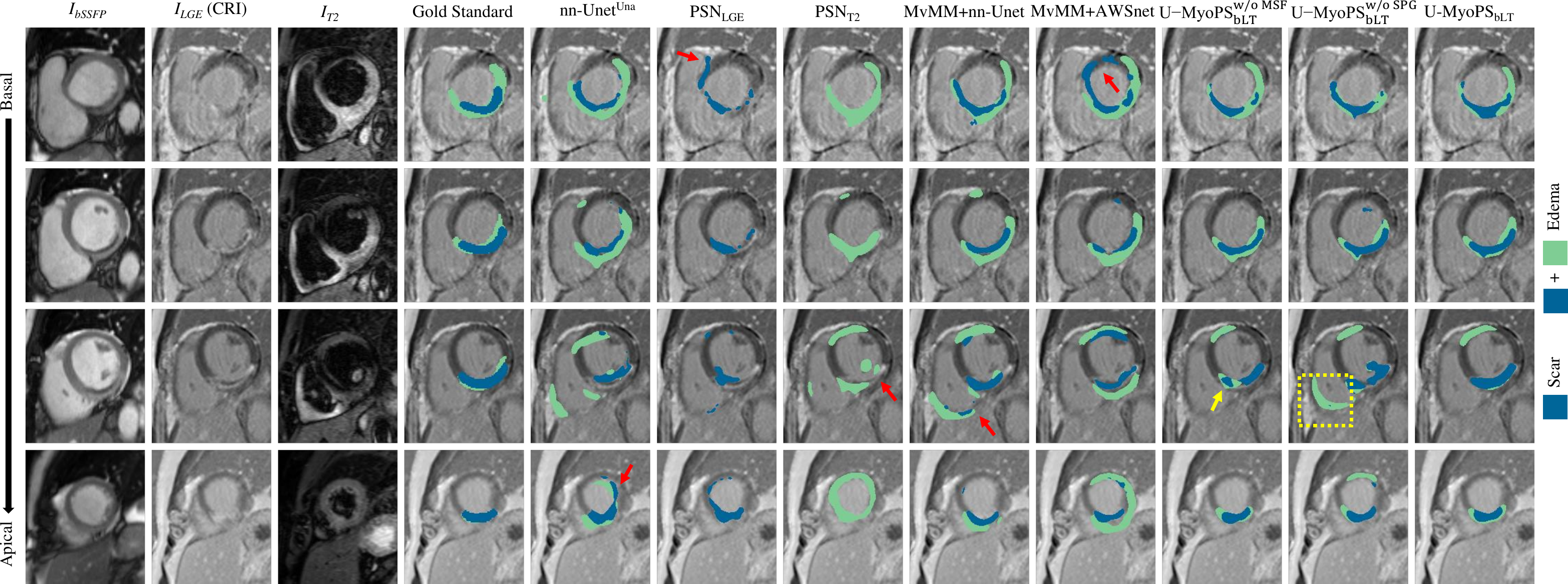}
		\caption{Visualization of MyoPS results from \hly{ basal to apical} slices. The results are overlaid on the \hly{LGE images}. The scarring and edema regions are colorized with blue and green, respectively. 
			Arrows, circles and rectangles mark the typical  regions where U-MyoPS$_{\text{\hly{bLT}}}$ achieved better results. \hly{ CRI (common reference image) indicates the selected reference image for multi-sequence registration.}
		}
		\label{fig:rj:U-MyoPS}    
	\end{figure*}

	The upper part of Table \ref{tab:rj:U-MyoPS} lists MyoPS results of different methods on pMM-CMR.
	\hly{MvMM+nn-Unet obtained better scar ($p<0.01$) and edema ($p<0.05$) results against nn-Unet$^{\text{Una}}$. The reason is that MvMM+nn-Unet utilized aligned \hly{MS-CMR images}, and the pathological and anatomical information from the aligned images could complement each other for segmentation. Meanwhile, U-MyoPS$_\text{bTL}$ achieved better results than nn-Unet$^{\text{Una}}$ for all evaluation metrics. One of the reasons is that  U-MyoPS$_\text{bTL}$ aligned MS-CMR for segmentation.} 
	
	\hly{U-MyoPS$_{\text{\hly{bLT}}}$ obtained better scar and edema results against mono-sequence semantic segmentation methods (\ie  PSN$_{\text{LGE}}$ and PSN$_{\text{T2}}$) which only utilized single sequence (LGE or T2) of CMR  images.}
	Particularly, \hly{for scar segmentation, U-MyoPS$_\text{bTL}$ improved Dice and HD by 8.92\% ($p<0.01$) and 13.84 mm ($p<0.01$) against PSN$_{\text{LGE}}$. Meanwhile, compared to PSN$_{\text{T2}}$, U-MyoPS$_\text{bTL}$ obtained better Dice (8.20\%, $p<0.01$) and HD (3.19 mm, $p=0.21$) results for edema segmentation.}
	This is because U-MyoPS$_{\text{\hly{bLT}}}$ aligned \hly{MS-CMR images} for \hly{MyoPS}. Segmentation methods could obtain more robust pixel-wise classification  based on the intensity information from aligned scarring and edema regions.

	\hly{U-MyoPS$_{\text{\hly{bLT}}}$ achieved better performance than the semantic segmentation method (\ie MvMM+nn-Unet and \hly{MvMM+AWSnet}) which consumed pre-aligned \hly{MS-CMR images}.} 
	One can see U-MyoPS$_{\text{\hly{bLT}}}$ outperformed MvMM+nn-Unet for scar ($p<0.01$) and edema ($p<0.01$)  segmentation results in terms of Dice and HD. The underlying reason was that U-MyoPS$_{\text{\hly{bLT}}}$ extracted \hly{myocardium} as spatial prior, facilitating the subsequent MyoPS performance. 
	\hly{Meanwhile, U-MyoPS$_{\text{\hly{bLT}}}$ obtained comparable results to MvMM+AWSnet on scar ($p=0.24$) and edema ($p=0.17$) segmentation in terms of Dice. 
		Nevertheless, U-MyoPS$_{\text{\hly{bLT}}}$ was computationally efficient. The average runtime of  MvMM+AWSnet was about 469 seconds (MvMM: 461 seconds; AWSnet: 8 seconds), whereas U-MyoPS$_{\text{\hly{bLT}}}$ only required about 10 seconds to segment pathologies from the unaligned \hly{MS-CMR} images of a subject. }  	
	
	\hly{Furthermore, we conducted an inter-observer variations study for MyoPS based on Dice metric.  The inter-observer variations were 70.4\% and 77.9\% for scar and edema segmentation in terms of Dice, respectively. As listed in Table \ref{tab:rj:U-MyoPS}, U-MyoPS$_\text{bLT}$ obtained 64.9\% and 76.0\% for scar and edema segmentation, respectively. 
		U-MyoPS$_\text{bLT}$ was worse than the inter-observer for scar segmentation, but close for edema segmentation.}

	\MyFig \ref{fig:rj:U-MyoPS} visualizes the MyoPS results of different methods.  Here, the gold standard pathology labels were the ones of U-MyoPS$_{\text{\hly{bLT}}}$\footnote{The scar and edema were manually delineated in original unaligned T2 and LGE images. U-MyoPS and MvMM+nn-Unet aligned these images into a common space, and performed MyoPS for the aligned images. Note that different methods may obtain different alignments of \hly{MS-CMR} images. Thus, the corresponding gold standard pathology label may also differ slightly.}. 
	\hly{As indicated by the red Arrows, U-MyoPS$_{\text{\hly{bLT}}}$ reduced inappropriate segmentation,  and mitigated outliers against \hly{comparison methods}. This demonstrated the promising of U-MyoPS$_{\text{\hly{bLT}}}$. 
	}

	\subsubsection{Ablation Study of MSF and SPG}\label{sec:exp:ablation:SPG}
	
	We implemented two variants of U-MyoPS$_{\text{\hly{bLT}}}$ for the ablation study of MSF and SPG: 
	\begin{itemize}
		\item U-MyoPS$^{\text{w/o MSF}}_{\text{\hly{bLT}}}$: Our U-MyoPS$_{\text{\hly{bLT}}}$ model without using MSF.
		
		\item U-MyoPS$^{\text{w/o SPG}}_{\text{\hly{bLT}}}$: Our U-MyoPS$_{\text{\hly{bLT}}}$ model without using SPG. 
	\end{itemize}

	The lower part of \hly{Table \ref{tab:rj:U-MyoPS} } presents the result of U-MyoPS$^{\text{w/o MSF}}_{\text{\hly{bLT}}}$. With MSF, U-MyoPS$_{\text{\hly{bLT}}}$ achieved slight better scar ($p=0.77$) and edema ($p=0.24$)  results to U-MyoPS$^{\text{w/o MSF}}_{\text{\hly{bLT}}}$  in term of Dice. Notably, as MSF was also proposed to improve the \hly{myocardium} structure extraction for MyoPS, one can refer to Section \ref{sec:exp:Myo:prior} for the ablation study of MSF on \hly{myocardium} structure extraction.
	
	Meanwhile, the lower part of \hly{Table \ref{tab:rj:U-MyoPS}} presents the result of  U-MyoPS$^{\text{w/o SPG}}_{\text{\hly{bLT}}}$. Without SPG, U-MyoPS$^{\text{w/o SPG}}_{\text{\hly{bLT}}}$ suffered performance degradation. For instance, compared to U-MyoPS$_{\text{\hly{bLT}}}$, U-MyoPS$^{\text{w/o SPG}}_{\text{\hly{bLT}}}$ decreased Sen by almost 4\% ($p<0.05$) and 5\% ($p=0.01$) for scar and edema segmentation, respectively. This is because the scarring and edema regions are in hly{myocardium}. It would turn out to be harder for U-MyoPS$^{\text{w/o SPG}}_{\text{\hly{bLT}}}$  to find more pathology regions without \hly{myocardium} prior information. 

	\MyFig \ref{fig:rj:U-MyoPS} visualizes the MyoPS results of  U-MyoPS$^{\text{w/o MSF}}_{\text{\hly{bLT}}}$ and U-MyoPS$^{\text{w/o SPG}}_{\text{\hly{bLT}}}$. Even U-MyoPS$^{\text{w/o MSF}}_{\text{\hly{bLT}}}$ obtained comparable quantity results  to U-MyoPS$_{\text{\hly{bLT}}}$, U-MyoPS$_{\text{\hly{bLT}}}$ still  achieved more reasonable details (see \hly{yellow} Arrows). The underlying reason is that U-MyoPS$_\text{\hly{bLT}}$ obtained more plausible \hly{myocardium} structures by using MSF (see Section \ref{sec:exp:Myo:prior}), which facilities segmentation details. 
	Moreover, by using SPG, U-MyoPS$_{\text{\hly{bLT}}}$ could reduce outliers (see \hly{yellow} Rectangles)  compared to U-MyoPS$^{\text{w/o SPG}}_{\text{\hly{bLT}}}$.  This indicated the benefit of employing spatial prior information for MyoPS. 

	\subsection{\hly{Results of Clinical Quantification}}\label{sec:seg:scar:eval}
	\hly{We evaluated the performance of U-MyoPS based on clinical indices, and compared U-MyoPS to clinical segmentation methods, namely 1-SD, 2-SD and 3-SD \cite{bondarenko2005standardizing}.} 
	\subsubsection{\hly{Results of scar size and transmurality}}
	\hly{ We employed two clinical indices, \ie scar size and transmurality, for scar quantification. Scar size was quantified as the percentage of LV myocardium. For transmurality, we adopted centerline chord method \cite{sheehan1986advantages,zhang2022artificial} to divide LV myocardium into 100 equally spaced chords. The transmurality of each chord was quantified as (scar pixels / chord pixels)$\times$100\%.}

\begin{figure}[htb] 
		\centering
		
		\includegraphics[width=1\textwidth]{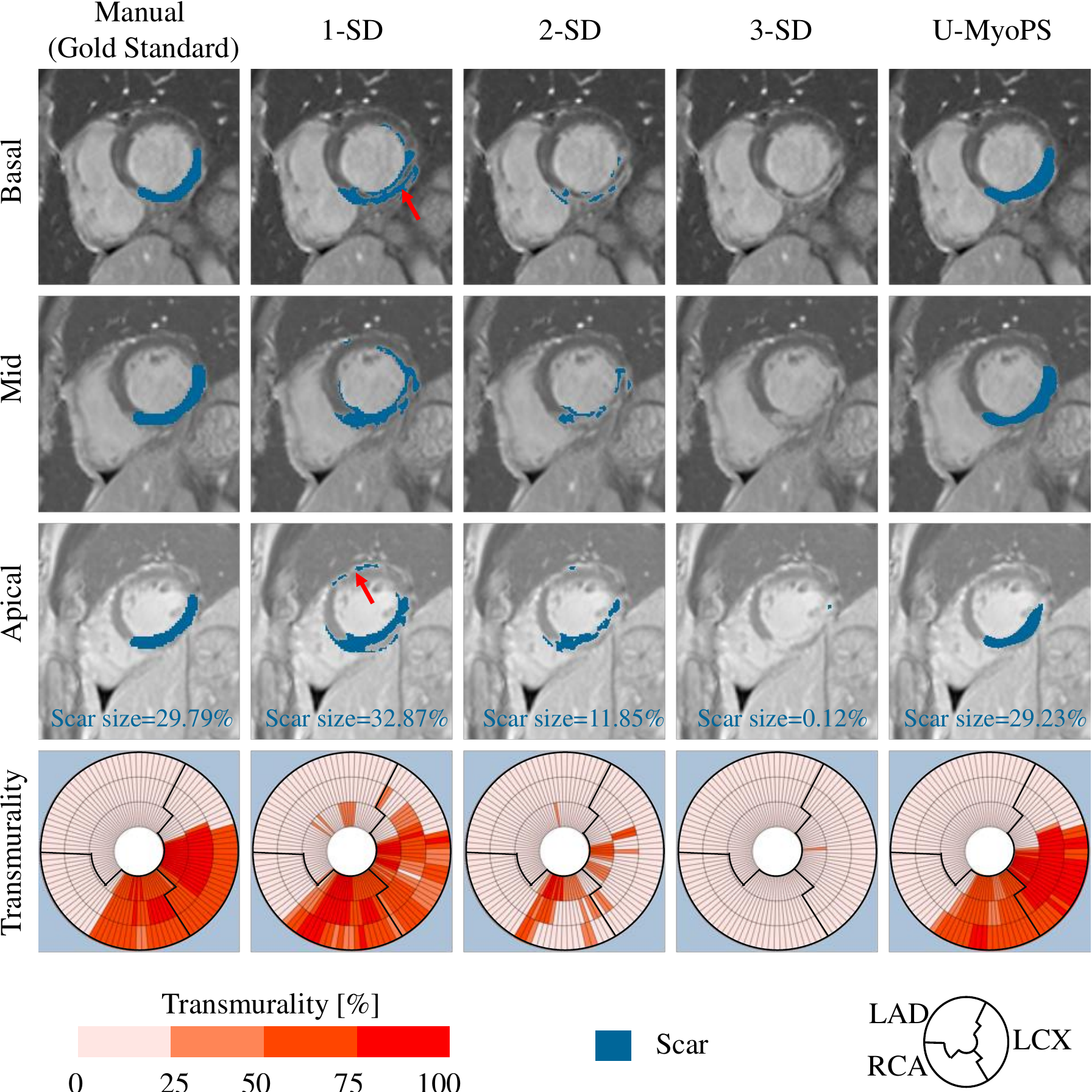}
		\caption{\hly{Examples of myocardial scar size and  transmurality.  The scar sizes as the percentage of LV myocardium are in the apical slices. The transmurality of each subject is demonstrated in the bull-eyes plots of the last low. 
				Each chord was encoded with different color according to its transmurality, \ie  mistyrose (viable): 0  $\sim$ 25\%; coral (likely viable): 26\% $\sim$ 50\%; orange red (likely nonviable) 51\%  $\sim$ 75\%; red (nonviable): 76\% $\sim$ 100\% \cite{zhang2022artificial}. Arrows indicate the regions where U-MyoPS achieved better results than 1-SD. LAD: left anterior descending artery; LCX: left circumflex artery;  RCA: right coronary artery}}
		\label{fig:zs:bulleye}    
	\end{figure}

	\hly{\MyFig \ref{fig:zs:bulleye} shows a typical scar on left circumflex territory.  1-SD and U-MyoPS obtained closely matching results (scar size and transmurality) to the manual delineation. Whereas, 2-SD and 3-SD suffered under-segmentation, resulting in inaccuracy for measurements. Moreover, n-SD methods are sensitive to noises, while U-MyoPS could generate more clear and plausible results as indicated by arrows. }
	
\begin{figure}[htp]
		\centering     
		\makebox[1\textwidth][c]{ \footnotesize {Scar size as the percentage of LV myocardium} }
		\subfigure{\label{fig:zs:infarct_size_nSD}\includegraphics[width=0.49\textwidth]{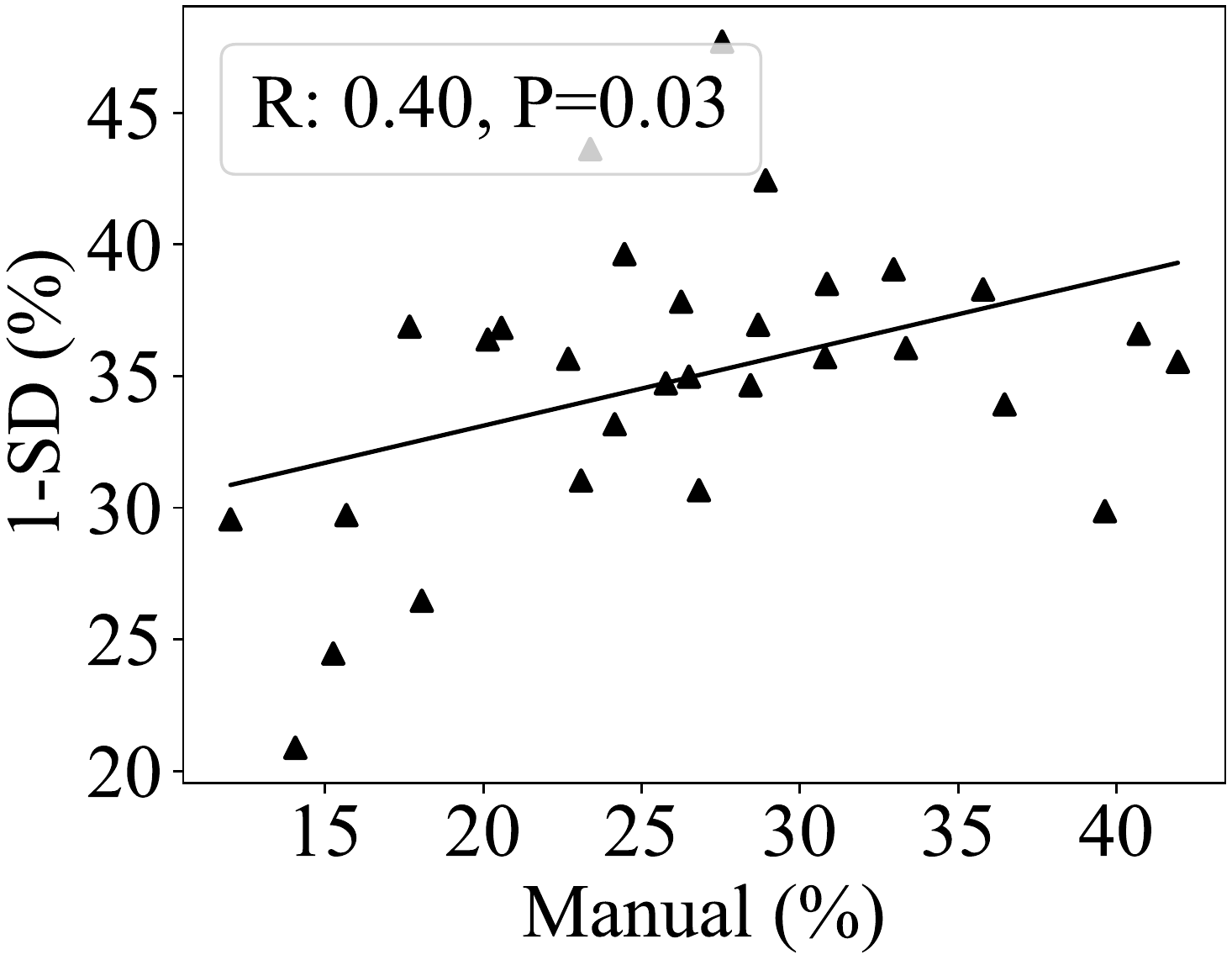}}
		\subfigure{\label{fig:zs:infarct_size_umyops}\includegraphics[width=0.49\textwidth]{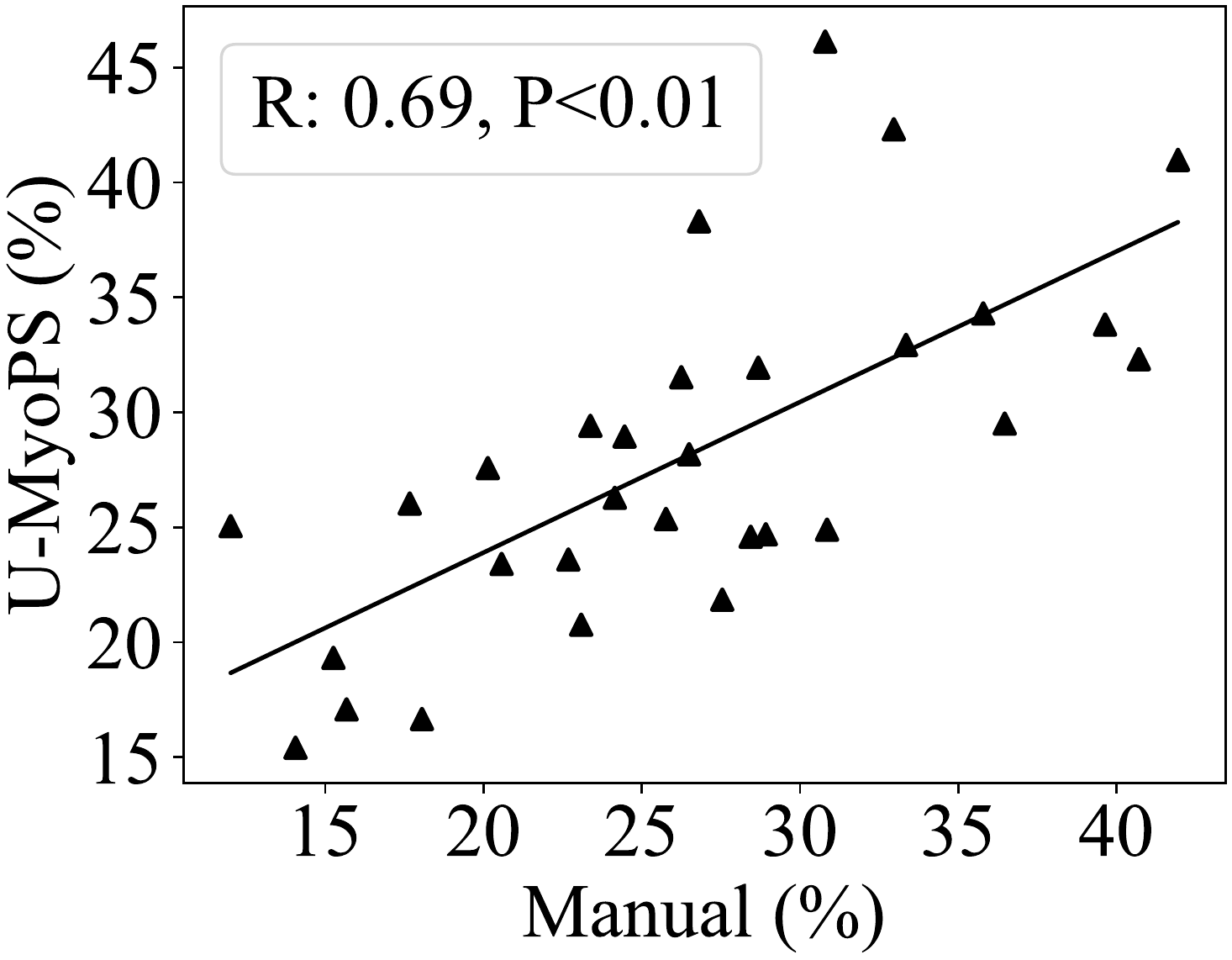}}
		\makebox[1\textwidth][c]{	\footnotesize {Number of transmural chords} }
		\subfigure{\label{fig:zs:transmurality_nSD}\includegraphics[width=0.49\textwidth]{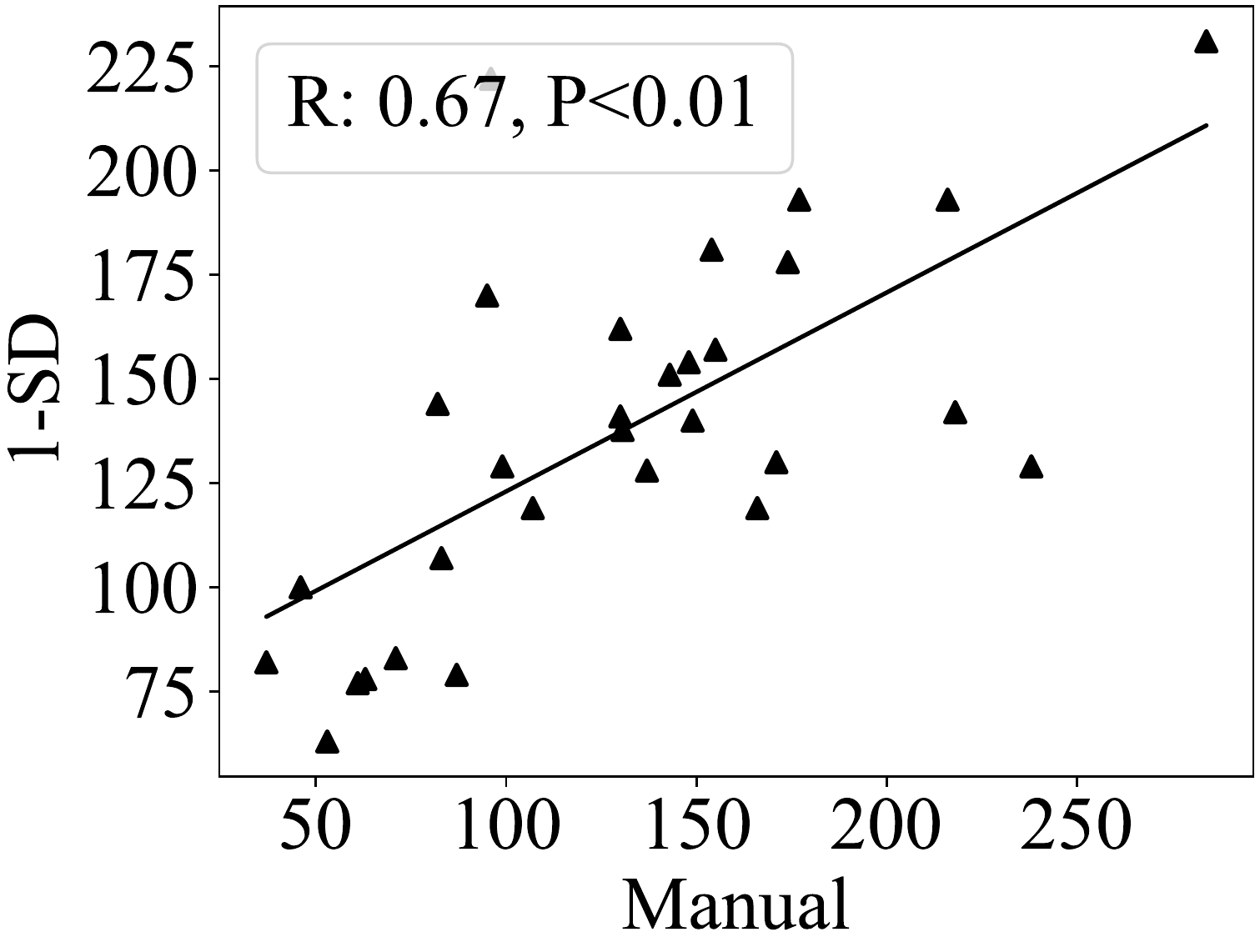}}
		\subfigure{\label{fig:zs:transmurality_umyops}\includegraphics[width=0.49\textwidth]{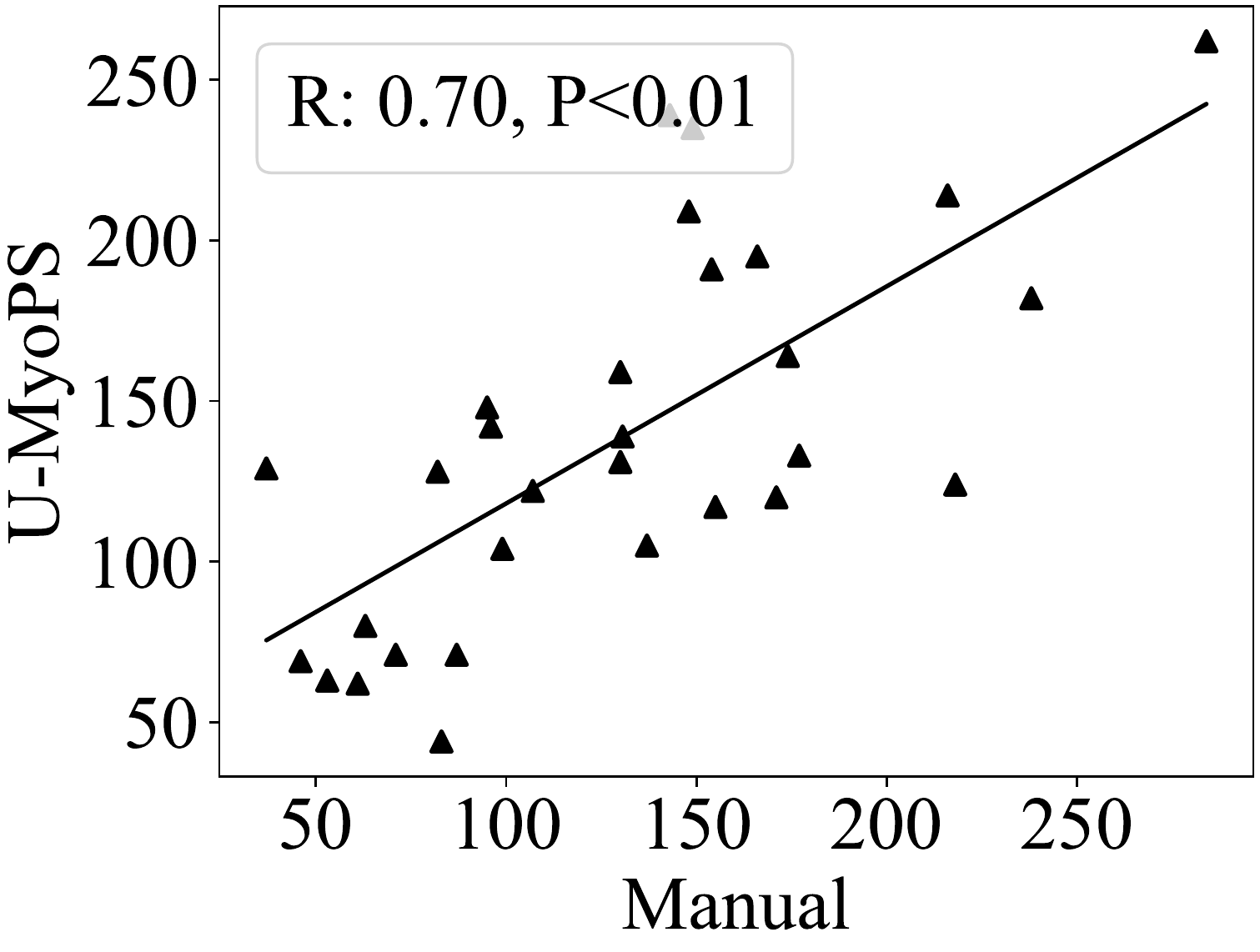}}
		\caption{\hly{Assessment of transmurality and scar size  of different methods. Here, we investigated the correlation between automatic methods (\ie 1-SD and U-MyoPS) and manual delineation. We quantified the transmurality of patients by counting the number of transmural chords (transmurality  $>$ 50\%) based on centerline chord method \cite{sheehan1986advantages,zhang2022artificial}. The top and bottom rows are correlation plots of scar size and transmurality  quantification  among different methods, respectively. R: Pearson correlation coefficients.} }
		\label{fig:zs:infarct_size} 
	\end{figure}

	\hly{
		Furthermore, we investigated the correlation among the quantification results of 1-SD\footnote{Note that 1-SD obtained optimal performance against 2-SD and 3-SD in pMM-CMR dataset, we thus only employed 1-SD methods for correlation study.}, U-MyoPS and manual delineation for each patient.  Compared to 1-SD, U-MyoPS obtained slightly better results for scar size quantification, and achieved comparable performance for transmural chord quantification  as shown in \MyFig \ref{fig:zs:infarct_size}. Meanwhile, U-MyoPS showed significant correlations with manual delineation, the Pearson correlation coefficients were  0.69 ($p<0.01$) and 0.70 ($p<0.01$) for scar size and transmural chord, respectively.  This supported that U-MyoPS could achieve robust performance for scar quantification.
	}

	\subsubsection{\hly{Results of edema size}}
\begin{figure}[ht] 
		\centering
		
		\includegraphics[width=1\textwidth]{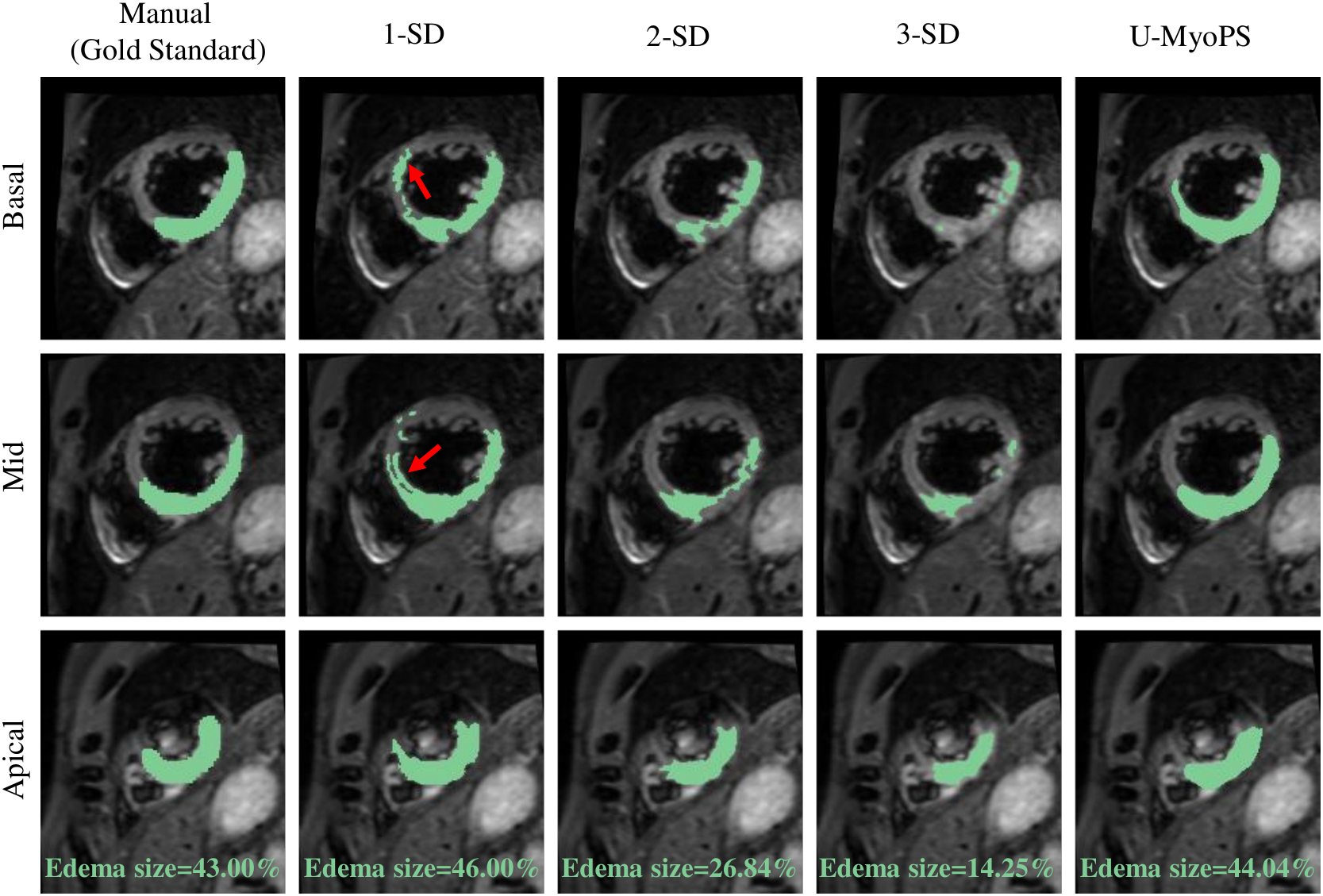}
		\caption{\hly{Examples of myocardial edema size.  The edema sizes as the percentage of LV myocardium are in the apical slices. Arrows indicate the regions where U-MyoPS obtained better results than 1-SD}}
		\label{fig:zs:edema_size}    
	\end{figure}

\begin{figure}[ht]
		\centering     
		\makebox[1\textwidth][c]{ \footnotesize {Edema size as the percentage of LV myocardium} }
		\subfigure{\label{fig:zs:edema_size_nSD}\includegraphics[width=0.49\textwidth]{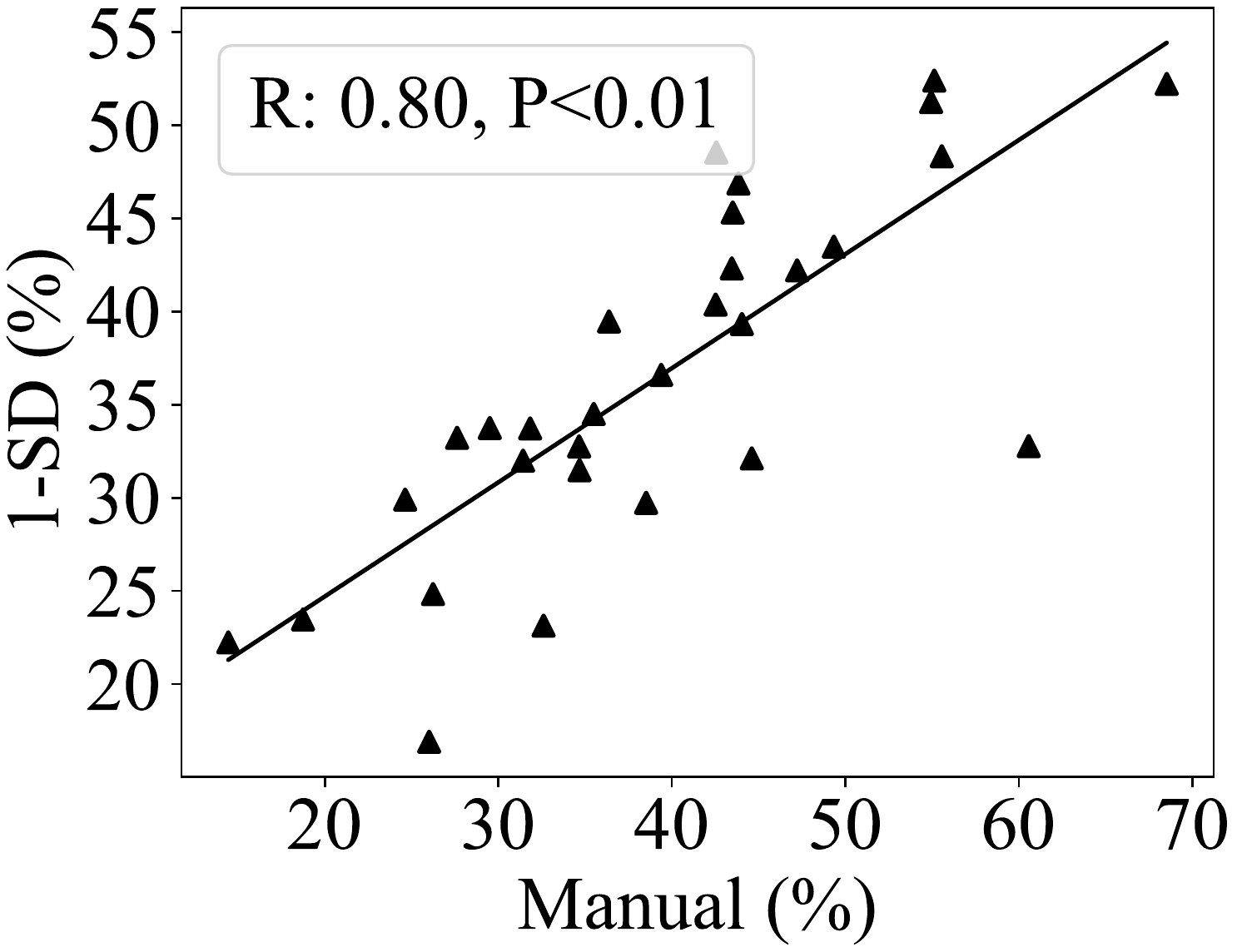}}
		\subfigure{\label{fig:zs:edema_size_umyops}\includegraphics[width=0.49\textwidth]{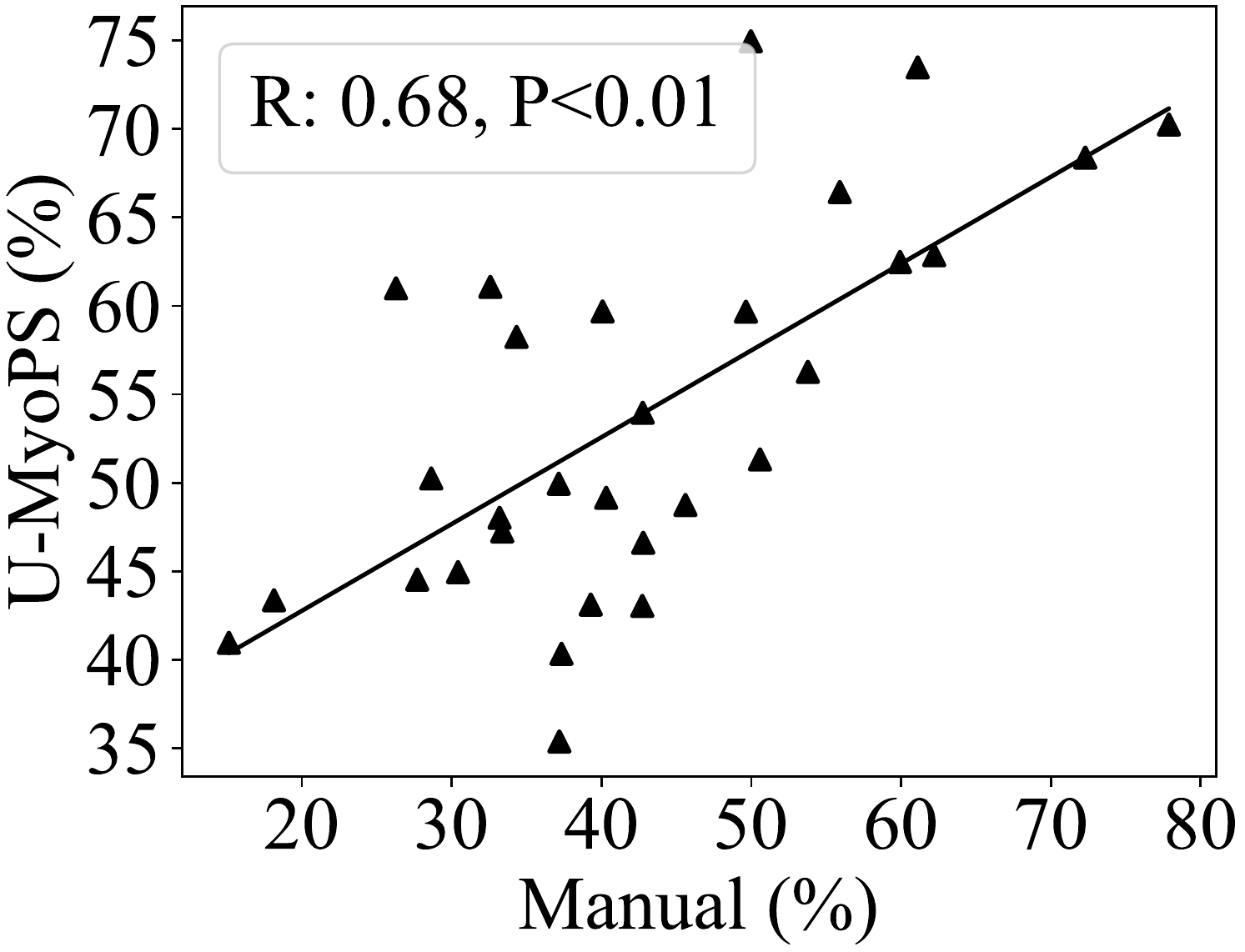}}
		\caption{\hly{Assessment of  edema size  of different methods. Here, we investigated the correlation between automatic methods (\ie 1-SD and U-MyoPS) and manual delineation. R: Pearson correlation coefficients.} }
		\label{fig:zs:edema_size_corr} 
	\end{figure}

	\hly{We quantified edema size as the percentage of LV myocardium. \MyFig \ref{fig:zs:edema_size} visualizes a typical edema. Both 1-SD and U-MyoPS obtained better segmentation results than 2-SD and 3-SD. Notably, U-MyoPS could obtain more robust results against 1-SD as indicated by arrows.  
		In \MyFig \ref{fig:zs:edema_size_corr}, we further investigated  the correlation among  1-SD, U-MyoPS and manual delineation. U-MyoPS obtained  worse Pearson correlation coefficients (0.80 \vs 0.68) against 1-SD. Nevertheless, U-MyoPS achieved a significant ($p<0.01$) correlation to manual delineation.}

	\subsection{Effects of the Number of Sequences} \label{exp:results:pMM-CMR:num of sequence}
	
	\begin{table*}[htb] 
		\resizebox{1\textwidth}{!}{ 
			\begin{tabular}{lcccc|cccc|cccc}
				\hline
				\multirow{2}{*}{Method} & \multicolumn{3}{c}{Sequences} & & \multicolumn{4}{|c}{Scar  }& \multicolumn{4}{|c}{Edema }\\ 
				
				& \hly{bSSFP}  & LGE  & T2 & NoS & Dice (\%) $\uparrow$  & Sen  (\%) $\uparrow$ & Pre (\%) $\uparrow$ & HD (mm)   $\downarrow$ & Dice (\%) $\uparrow$  & Sen  (\%) $\uparrow$ & Pre (\%) $\uparrow$ & HD (mm)   $\downarrow$ \\
				\hline 
				
				U-MyoPS$_\text{\hly{bL}}$ & \checkmark & \checkmark & $\times$ & 2 &  61.72 (11.48) & 63.95 (13.65) & {60.85 (12.31)} & 35.26 (16.63)  &  \multicolumn{4}{c}{N/A} \\	
				
				U-MyoPS$_{\text{LT}}$   & $\times$ & \checkmark & \checkmark & 2 & 62.32 (10.58) & 61.56 (12.26) & \textbf{64.61 (12.21)} & 30.44 (17.58) &{74.80 (12.29)} & {79.86 (10.62)} & 71.48 (15.47) & {29.18 (20.17)}\\

				U-MyoPS$_{\text{\hly{bLT}}}$ & \checkmark & \checkmark & \checkmark & 3 &  \textbf{64.92 (9.816)}   & \textbf{68.30 (12.56) } & {63.34 (11.71)} & \textbf{29.16 (16.65)} & \textbf{76.01 (9.784)} & \textbf{80.49 (8.942)} & \textbf{73.53 (14.05)} & \textbf{27.89 (18.45)} \\
				
				\hline
				
			\end{tabular}
			
		}{
			\caption{Performance of  U-MyoPS with different numbers of CMR sequence images.   
				Subscript ``$_\text{\hly{bL}}$'': \hly{bSSFP} and LGE images; 
				Subscript ``$_\text{LT}$'': LGE and T2 images; Subscript ``$_\text{\hly{bLT}}$'': \hly{bSSFP}, LGE and T2 images.  NoS: Number of sequences. The best results are labeled in \textbf{bold}.}
			\label{tab:rj:multi_seq}
		}
		
	\end{table*}	
	
	U-MyoPS aligned \hly{MS-CMR images} for MyoPS. \hly{Since each of the sequences could provide different information, the myocardium and pathology segmentation could be improved by fusing more sequences.}   We further evaluated the effects of aligning different numbers of sequences on pMM-CMR dataset. 
	
	\subsubsection{MyoPS \vs Number of Sequences}
	We first investigated the  performance of U-MyoPS with different numbers of sequences. 
	Here, we implemented additional \hly{two} variants of U-MyoPS with different numbers of sequences:
	\begin{itemize}
		
		\item U-MyoPS$_\text{\hly{bL}}$: U-MyoPS which utilized $\{I_{\hly{bSSFP}}, I_{LGE} \}$ images for \textit{scar} segmentation.
		
		\item U-MyoPS$_{\text{LT}}$: U-MyoPS which utilized $\{I_{LGE}, I_{T2}\}$ images for \textit{scar and edema}  segmentation.
	\end{itemize}
	
	Table \ref{tab:rj:multi_seq} lists the MyoPS performance of U-MyoPS framework with different numbers of CMR sequences. U-MyoPS was capable of processing CMR images with different numbers (two or three) of sequences. Generally, the method using three sequences, \ie U-MyoPS$_{\text{\hly{bLT}}}$, achieved the best results in most metrics.
	Even the method using two sequences, \ie U-MyoPS$_{\text{LT}}$, obtained best Pre of scar, the difference ($p=0.56$) was not obvious when compared to U-MyoPS$_{\text{\hly{bLT}}}$.
	Moreover, U-MyoPS$_{\text{\hly{bLT}}}$ increased the Sen of scars by about 6\% ($p<0.01$) and 3\% ($p=0.05$) with additional sequence images when compared to U-MyoPS$_\text{LT}$ and U-MyoPS$_\text{\hly{bL}}$, respectively.  This implied the potential advantages of aligning more sequences for MyoPS.
	
	\subsubsection{\hly{Myocardium} Extraction \vs Number of Sequences}
	\label{sec:exp:Myo:prior}
	
	We investigated the performance of \hly{myocardium} extraction with different numbers of sequences. 
	Here, we presented three \hly{myocardium} extraction methods with different numbers of sequences: 
	\begin{itemize}
		
		\item $\text{Myo}_\text{\hly{bL}}$ and $\text{Myo}_\text{LT}$: The anatomical structure extraction method of  (see Section \ref{sec:prior}) U-MyoPS which fused $\{I_{\hly{bSSFP}}, I_{LGE}\}$ and $\{I_{LGE}, I_{T2}\}$ for \hly{myocardium}  extraction, respectively. 
		
		\item  $\text{Myo}_\text{\hly{bLT}}$: The anatomical structure extraction method of U-MyoPS that  fused  $\{I_{\hly{bSSFP}}, I_{LGE}, I_{T2}\}$ for \hly{myocardium} extraction.
		
	\end{itemize}
	Moreover, we implemented a variant of $\text{Myo}^{\text{}}_{\text{\hly{bLT}}}$ to investigate the effectiveness of the proposed fuse schema, \ie MSF, on \hly{myocardium} extraction.
	
	\begin{itemize}
		\item  $\text{Myo}^{\text{w/o MSF}}_{\text{\hly{bLT}}}$:  $\text{Myo}_\text{\hly{bLT}}$ which extracted \hly{myocardium} of CRIs without fusing multi-sequence information via MSF.
		
	\end{itemize}

	%
	%

	\begin{table}[htb] 
		
		\resizebox{1\textwidth}{!}{ 
			\begin{tabular}{lcccc|ccc}
				\hline

				\hline
				\multirow{2}{*}{Method} & \multicolumn{3}{c}{Sequences} & & \multicolumn{2}{c}{Myocardium}& \\ 
				
				& \hly{bSSFP}  & LGE  & T2  & NoS & {Dice (\%)  $\uparrow$} &{HD (mm)  $\downarrow$} \\
				
				\hline 

				
				$\text{Myo}_\text{\hly{bL}}$   &\checkmark & \checkmark & $\times$ & 2 &  {83.39 (4.829)  }&{18.23 (21.18)}\\
				
				$\text{Myo}_\text{LT}$  & $\times$ & \checkmark & \checkmark   & 2 &83.72 (3.334) & 16.15 (18.35)     \\
				
				$\text{Myo}_\text{\hly{bLT}}$ &  \checkmark & \checkmark & \checkmark & 3 &  \textbf{85.87 (2.786)} & \textbf{9.238 (7.872)}\\
				
				\hline
				\hline
				$\text{Myo}^{\text{w/o MSF}}_{\text{\hly{bLT}}}$ & \checkmark & \checkmark & \checkmark & 3 & {84.13 (4.544)}  &  12.88 (16.86) \\
				
				\hline

			\end{tabular}
		}
		\caption{Performance of the different \hly{myocardium} extraction methods.
			Subscript ``$_\text{\hly{bL}}$'': \hly{bSSFP} and LGE images; Subscript ``$_\text{LT}$'': LGE and T2 images; Subscript ``$_\text{\hly{bLT}}$'': \hly{bSSFP}, LGE and T2 images. NoS: Number of sequences.	The best results are labeled in \textbf{bold}.
		} 
		\label{tab:rj:prior}
		
	\end{table}

\begin{figure}[htp] 
		\centering
		\includegraphics[width=1\textwidth]{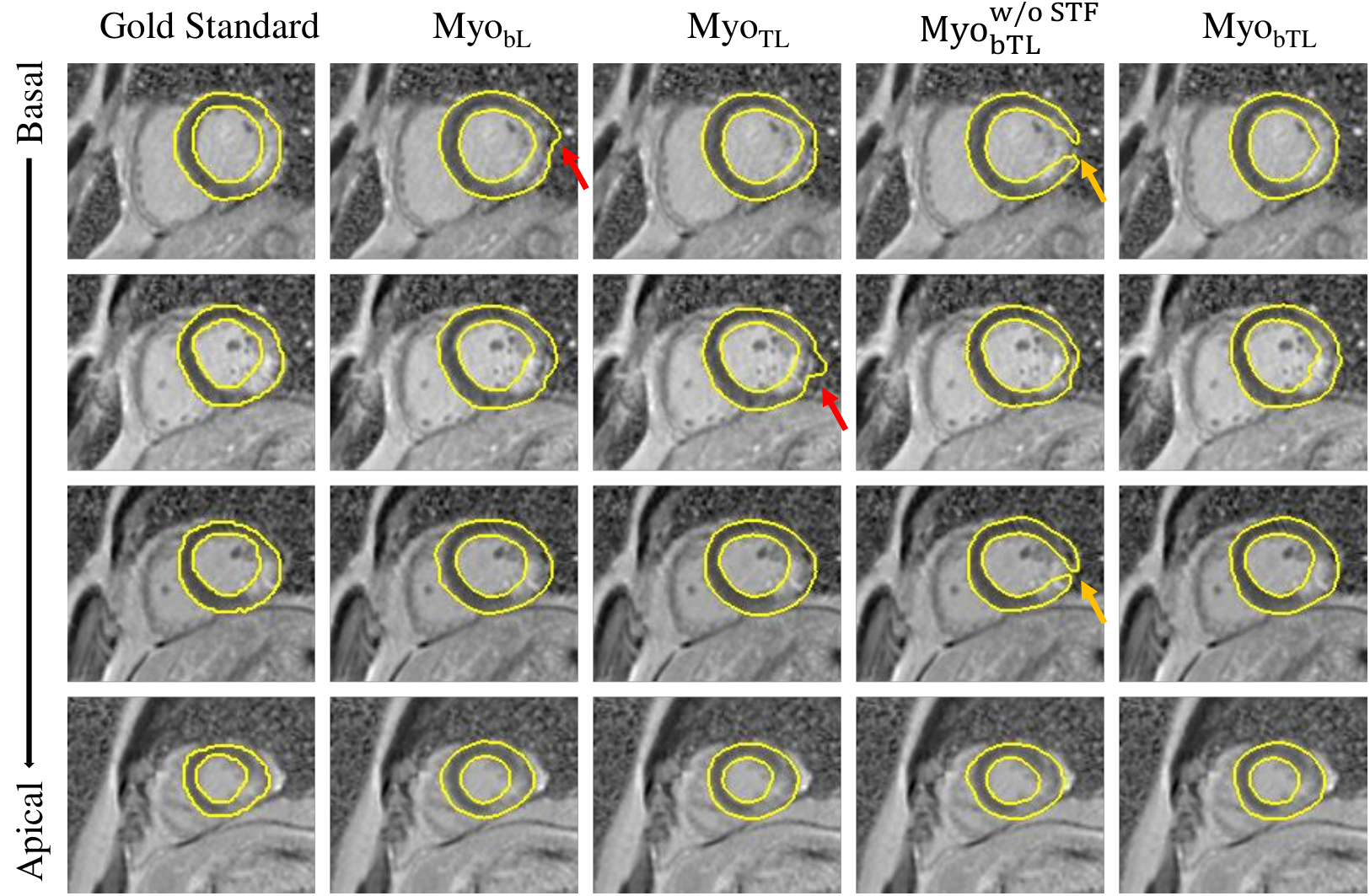}
		\caption{Visualization of \hly{myocardium} extraction results from \hly{basal to apical} slices. The results are overlaid on common reference images (CRIs). 
			Yellow curves indicate the predicted \hly{myocardium} and gold standard \hly{myocardium contours}  of \hly{CRIs}.  Arrows point to myocardial pathology regions which could easily degrade \hly{myocardium} extraction methods.}
		\label{fig:rj:prior}    
	\end{figure}

	Table \ref{tab:rj:prior} lists the results of different \hly{myocardium}  extraction methods. 
	Overall, the method using three sequences (\ie $\text{Myo}_\text{\hly{bLT}}$) outperformed the methods using two sequences (\ie $\text{Myo}_\text{\hly{bL}}$ and $\text{Myo}_\text{LT}$). For example, $\text{Myo}_\text{\hly{bLT}}$ improved the Dice and HD by almost 2\% ($p<0.01$) and 7 mm ($p=0.07$) against $\text{Myo}_\text{LT}$, respectively. This demonstrated the advantage of aligning more sequences of images for \hly{myocardium}  extraction.
	Meanwhile, \MyFig \ref{fig:rj:prior} visualizes different  \hly{myocardium} extraction results. One can observe  the myocardial pathology regions  easily degraded the \hly{myocardium} extraction performance (see red Arrows). Whereas, $\text{Myo}^{\text{}}_{\text{\hly{bLT}}}$ obtained more plausible details than other methods by using three CMR sequences.

	Furthermore, we investigated the effectiveness of MSF for \hly{myocardium} extraction.  Even $\text{Myo}^{\text{w/o MSF}}_{\text{\hly{bLT}}}$ consumed three sequence images, it did not explicitly fuse multi-sequence feature maps for \hly{myocardium}  extraction. Based on MSF, $\text{Myo}_\text{\hly{bLT}}$ could increased Dice by almost 2\% ($p<0.01$) against $\text{Myo}^{\text{w/o MSF}}_{\text{\hly{bLT}}}$. In \MyFig \ref{fig:rj:prior}, one can also observe that $\text{Myo}_\text{\hly{bLT}}$ achieved robust results than  $\text{Myo}^{\text{w/o MSF}}_{\text{\hly{bLT}}}$ (see orange Arrows). 
	
\begin{figure}[htb] 
		\centering
		\includegraphics[width=1\textwidth]{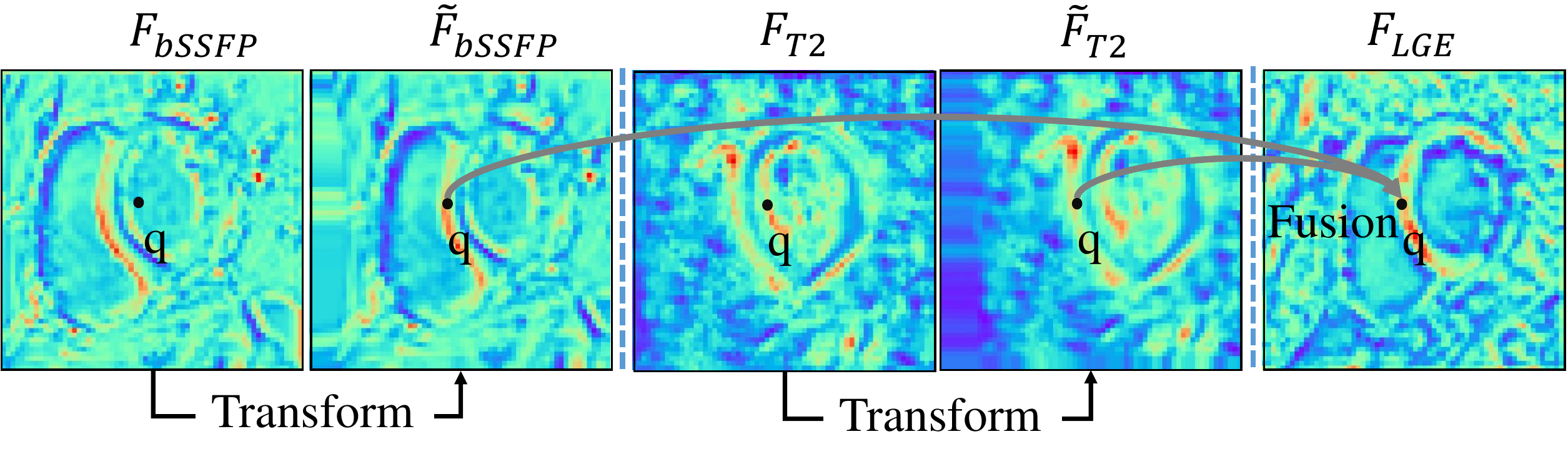}
		\caption{Illustration of fusing the feature maps of multi-sequence cardiac magnetic resonance images via MSF.  ${{F}}_{\hly{bSSFP}}$, ${{F}}_{T2}$ and ${{F}}_{LGE}$ are the feature maps from  3-rd level of $E_{\hly{bSSFP}}$, $E_{T2}$ and $E_{LGE}^{}$, respectively. $\tilde{{F}}_{\hly{bSSFP}}$ and $\tilde{{F}}_{T2}$ are the spatially transformed of ${{F}}_{\hly{bSSFP}}$ and ${{F}}_{T2}$, respectively. Each ${q}$ represents a point with a fixed spatial coordinate in feature maps. The semantic information of  feature maps (\ie $\tilde{{F}}^{}_{\hly{bSSFP}}$, $\tilde{{F}}^{}_{T2}$ and ${{F}}^{}_{LGE}$) in ${q}$ could become consistent after transformation. }
		\label{fig:rj:feat_MSF}    
	\end{figure}

	Additionally, we visualized the procedure of fusing misaligned feature maps of MSF in \MyFig \ref{fig:rj:feat_MSF}. We first showed the corresponding feature maps (${{F}}_{\hly{bSSFP}}$, ${{F}}_{T2}$ and ${F}_{LGE}$) of unaligned \hly{MS-CMR images} (${{I}}_{\hly{bSSFP}}$, ${{I}}_{T2}$ and ${I}_{LGE}$). For better demonstration, we placed a fixed spatial point $q$ on each feature map. Initially, the semantic information of  ${{F}}_{\hly{bSSFP}}$,  ${{F}}_{T2}$ and ${F}_{LGE}$ in $q$ could be incorrectly integrated via existing fusion operations, such as channel-wise concatenation \cite{li2020dual} and max-fusion \cite{jiang2020max}. 
	Whereas MSF could fuse ${F}_{\hly{bSSFP}}$ (or ${{F}}_{T2}$)  with ${F}_{LGE}$ by transforming ${F}_{\hly{bSSFP}}$  (or ${{F}}_{T2}$) into ${\tilde{F}}_{\hly{bSSFP}}$ (or ${\tilde{F}}_{T2}$). 
	Therefore, MSF provided a more appropriate and reliable way to fuse multi-sequence information.
	
	\subsection{\hly{Results of Multi-Sequence Registration on pMM-CMR Dataset}}\label{sec:reg:res}
	\hly{We investigated the registration performance of U-MyoPS by comparing it with three  state-of-the-art registration methods\footnote{ANTs:~{https://github.com/ANTsX/ANTsPy};\\ 
			VoxelMorph:~{https://github.com/voxelmorph/voxelmorph};\\
			RegNet$_\text{mvmm}$:~{https://github.com/xzluo97/MvMM-RegNet}}.
		\begin{itemize}
			\item ANTs \cite{avants2009advanced}: One of the state-of-the-art conventional affine + deformable registration tools. We invoked \textit{``SyN''} {registration} function based on their Python implementation. 
			\item VoxelMorph\cite{balakrishnan2019voxelmorph}: One of the state-of-the art networks which could perform pair-wise registration. The network was mainly trained by Dice loss. 
			\item RegNet$_\text{mvmm}$ \cite{luo2020mvmm}: The multi-sequence registration network which could jointly register \hly{MS-CMR images}. The network was mainly trained based on MvMM loss function. 
			\item U-MyoPS$_\text{Reg}$: The multi-sequence registration module of U-MyoPS. 
		\end{itemize}
	}
	\hly{Moreover, we implemented a variants of U-MyoPS$_\text{Reg}$ to investigated the effectiveness of constraint loss [see \eqref{eq:cons}].
		\begin{itemize}
			\item U-MyoPS$_\text{Reg}^\text{w/o Cons}$: U-MyoPS$_\text{Reg}$ without utilizing constraint loss during network training.
		\end{itemize}
	}
	
	\hly{Table \ref{tab:reg_res} summaries the registration results of different methods. One can observe that all registration methods could mitigate the misalignment among bSSFP, T2 and LGE images. Effectively, U-MyoPS$_\text{Reg}$ obtained the best Dice and HD results among all the registration methods. Particularly,  U-MyoPS$_\text{Reg}$ obtained better Dice (6.2\%, $p<0.05$) and HD (9.4 mm, $p<0.01$) performances than ANTs for T2 to LGE registration. Meanwhile, 
		compared to RegNet$_\text{mvmm}$, U-MyoPS$_\text{Reg}$ improved the Dice  of bSSFP to LGE and T2 to LGE registration by 7.8\% ($p<0.01$) and 5.2\% ($p<0.01$), respectively.
		Furthermore,  compared to the state-of-the-art registration network Voxelmorph, U-MyoPS$_\text{Reg}$ obtained 2.4\% ($p<0.05$)  and 2.7\%  ($p<0.05$)  Dice improvements for bSSFP to LGE and T2 to LGE registration, respectively.  This indicated that U-MyoPS obtained promising performance for registration. }
	
	\hly{Moreover, the lower part of Table \ref{tab:reg_res}  presents the result of U-MyoPS$_\text{Reg}^\text{w/o Cons}$. Without constraint loss [see \eqref{eq:cons}],  U-MyoPS$_\text{Reg}^\text{w/o Cons}$ suffered performance degradation for registration. For instance,  U-MyoPS$_\text{Reg}^\text{w/o Cons}$ decreased Dice (2.5\%, $p<0.01$)  and HD (0.65 mm, $p=0.12$) for bSSFP to LGE registration. This implied the benefit of constraint loss.
	}
	
	\begin{table}[htp]
		\centering
		\caption{\hly{The registration accuracy of different methods on myocardium. bSSFP$\rightarrow$LGE: bSSFP to LGE registration; T2$\rightarrow$LGE: T2 to LGE registration.}}
		\resizebox{\textwidth}{!}{
			\begin{tabular}{l|llll}
				\hline
				\multicolumn{1}{c|}{\multirow{2}{*}{\hly{Method}}} & \multicolumn{2}{c}{\hly{bSSFP$\rightarrow$LGE}}                 & \multicolumn{2}{c}{\hly{T2$\rightarrow$LGE}}                    \\ 
				\multicolumn{1}{c|}{}                        & \multicolumn{1}{c}{\hly{Dice (\%) $\uparrow$}} & \multicolumn{1}{c}{\hly{HD (mm) $\downarrow$}} & \multicolumn{1}{c}{\hly{Dice (\%) $\uparrow$}} & \multicolumn{1}{c}{\hly{HD (mm) $\downarrow$}} \\ \hline
				\hly{Initial}                                       & \hly{51.53 (18.01)}               & \hly{35.34 (40.34)}               & \hly{39.98 (18.04)}               & \hly{71.10 (98.52)}               \\
				\hdashline
				\hly{ANTs}                                         & \hly{78.44 (6.060)}                & \hly{10.80 (4.045)}                & \hly{73.11 (13.46) }              & \hly{18.85 (9.534)  }             \\
				\hly{RegNet$_\text{mvmm}$}          &  \hly{71.21 (11.80)}       & \hly{12.41 (4.542) }                            &     \hly{74.05 (10.36)}  & \hly{11.79 (4.930) }                            \\
				\hly{Voxelmorph}                                   & \hly{76.63 (9.756)}               & \hly{12.89 (7.332)}               & \hly{76.55 (7.470)}                & \hly{14.45 (7.975) }              \\
				\hly{U-MyoPS$_\text{Reg}$}                              &\hly{\textbf{79.07 (5.060)}}                & \hly{\textbf{8.548 (2.001)} }              &\hly{\textbf{79.26 (4.824)}}               & \hly{\textbf{9.427 (3.398)}}               \\ \hline
				\hline
				\hly{U-MyoPS$_\text{Reg}^\text{w/o Cons}$}    &   \hly{76.57 (5.893)}            &   \hly{9.200 (2.354) }           &  \hly{77.95 (6.663) }              &\hly{ 9.816 (3.974)}               \\
				\hline
			\end{tabular}
		}
		\label{tab:reg_res}
	\end{table}

	\subsection{Correlation Study}\label{exp:results:pMM-CMR:correlation}

	U-MyoPS performed multi-sequence registration, \hly{myocardium} extraction and MyoPS tasks. 
	\hly{The multi-sequence registration and myocardium extraction could promote  MyoPS task (see Section \ref{sec:compare_existing_method}).}
	We further investigated \hly{the} correlations among these tasks \hly{for} U-MyoPS$_\text{\hly{bLT}}$, and measured the correlations by calculating coefficient of determination (R$^2$). 
	
	\begin{figure}[htp]
		\centering     
		\subfigure{\label{fig:corr:reg_myops:a}\includegraphics[width=0.49\textwidth]{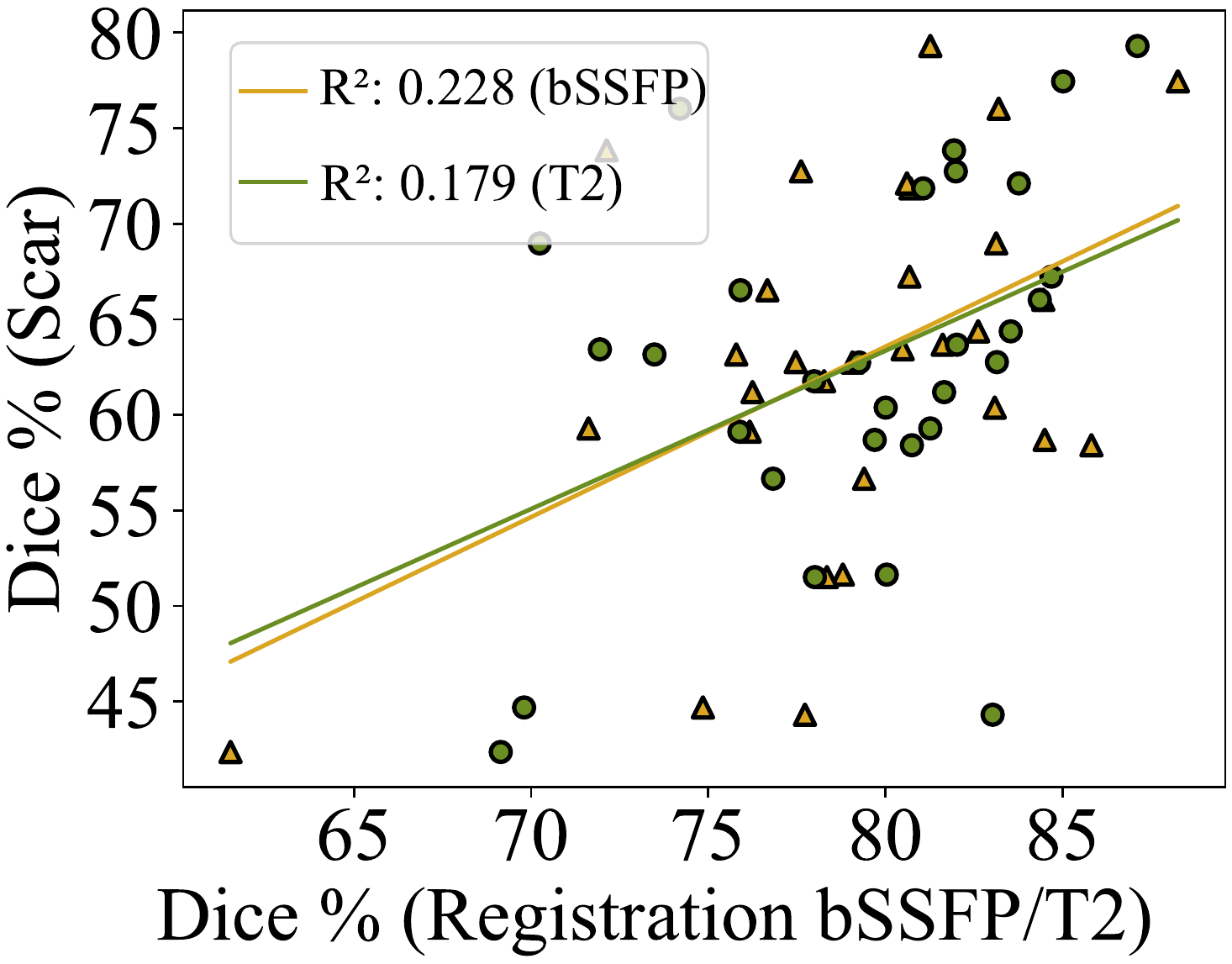}}
		\subfigure{\label{fig:corr:reg_myops:b}\includegraphics[width=0.49\textwidth]{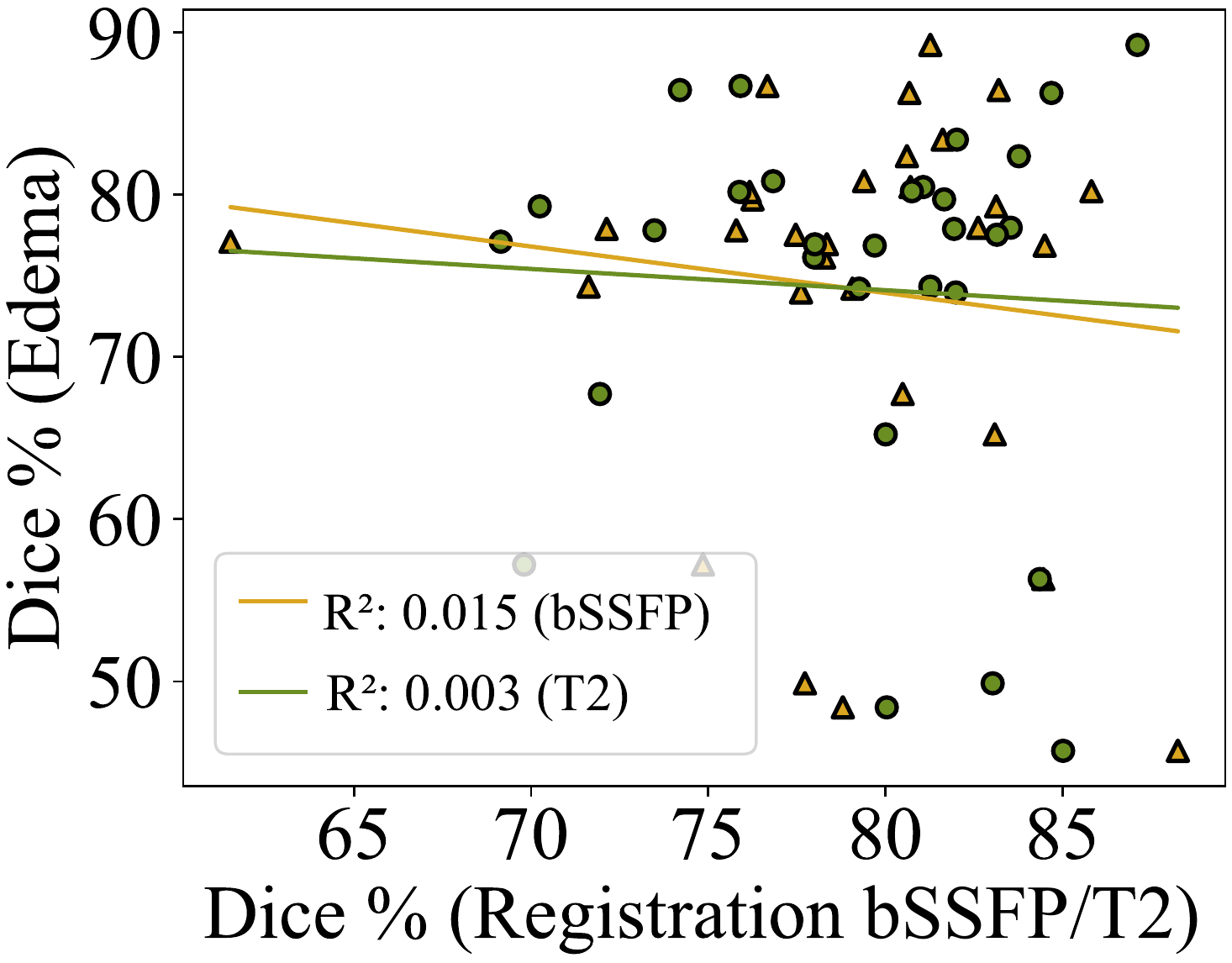}}
		\caption{Correlation between the accuracy of multi-sequence registration and MyoPS.}
		\label{fig:corr:reg_myops} 
	\end{figure}

	\subsubsection{Correlation between Multi-sequence Registration and MyoPS }\label{sec:corre:reg_myops}
	
	$R_{\hly{bSSFP}}$ (or $R_{T2}$) predicted TPS transformation parameters to align \hly{bSSFP} (or T2) to LGE images, and we measured the registration accuracy via the Dice between the \hly{myocardium} label of the warped \hly{bSSFP} (or T2) and LGE images. The left of \MyFig \ref{fig:corr:reg_myops} shows that the registration accuracy of \hly{bSSFP} and T2 had a correlation to the scar segmentation. 
	This is because that both the boundary and edema information in the warped \hly{bSSFP} and T2 images would complement the scar segmentation. Fusing more accurate information could further improve the corresponding scar segmentation.
	\hly{Note that} there was an insignificant relationship between the registration accuracy and edema segmentation as shown in the right of \MyFig \ref{fig:corr:reg_myops}. \hly{This is probably because the multi-sequence registration of U-MyoPS has already achieved sufficient accuracy to fuse complementary and relevant information for edema segmentation.}

	\begin{figure}[htb] 
		\RawFloats
		\begin{minipage}[t]{.49\linewidth}
			\includegraphics[width=\linewidth]{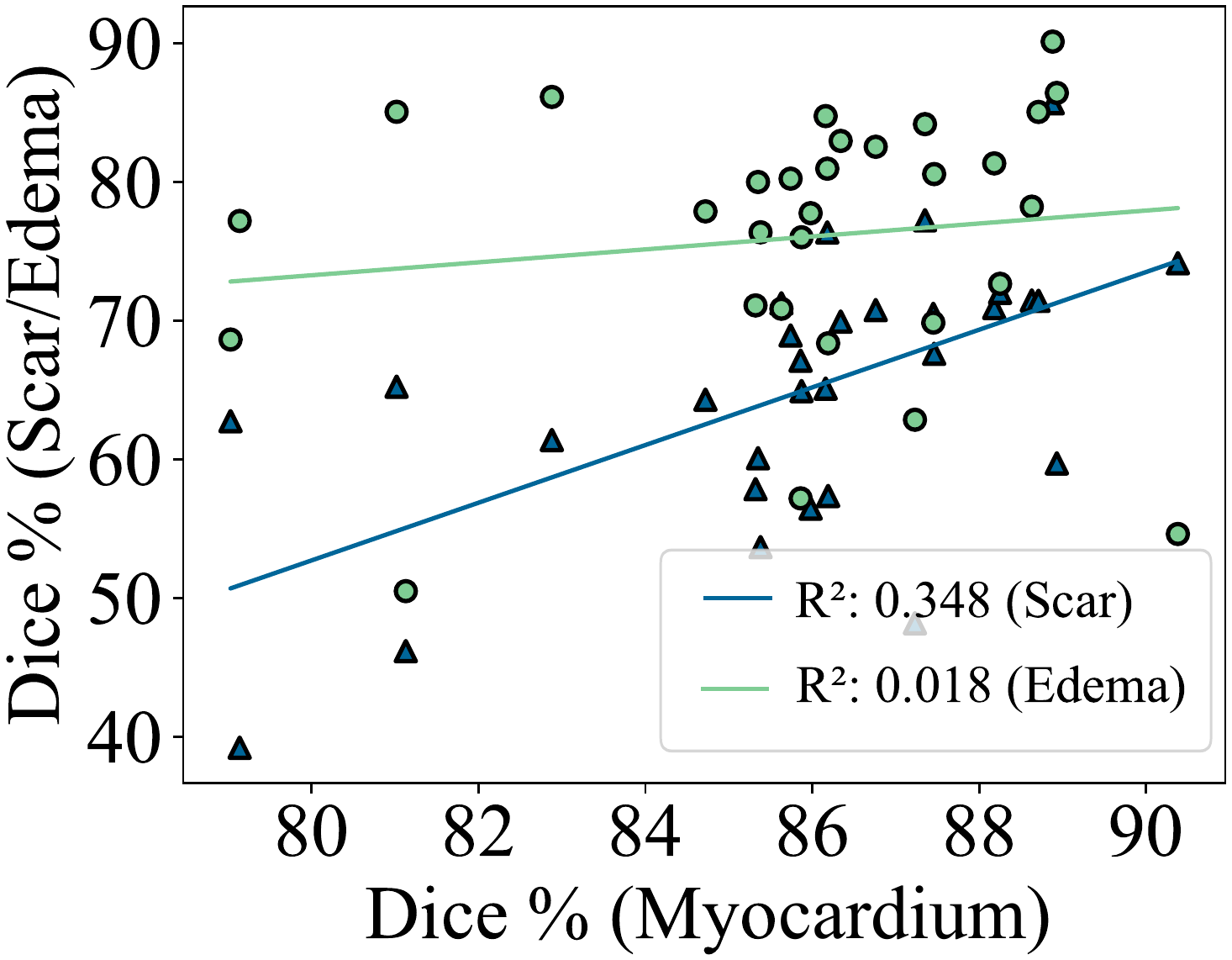}
			\caption{Correlation between the accuracy of \hly{myocardium} extraction and MyoPS .}
			\label{fig:corr:prior_myops}
		\end{minipage}
		\begin{minipage}[t]{.49\linewidth}
			\includegraphics[width=\linewidth]{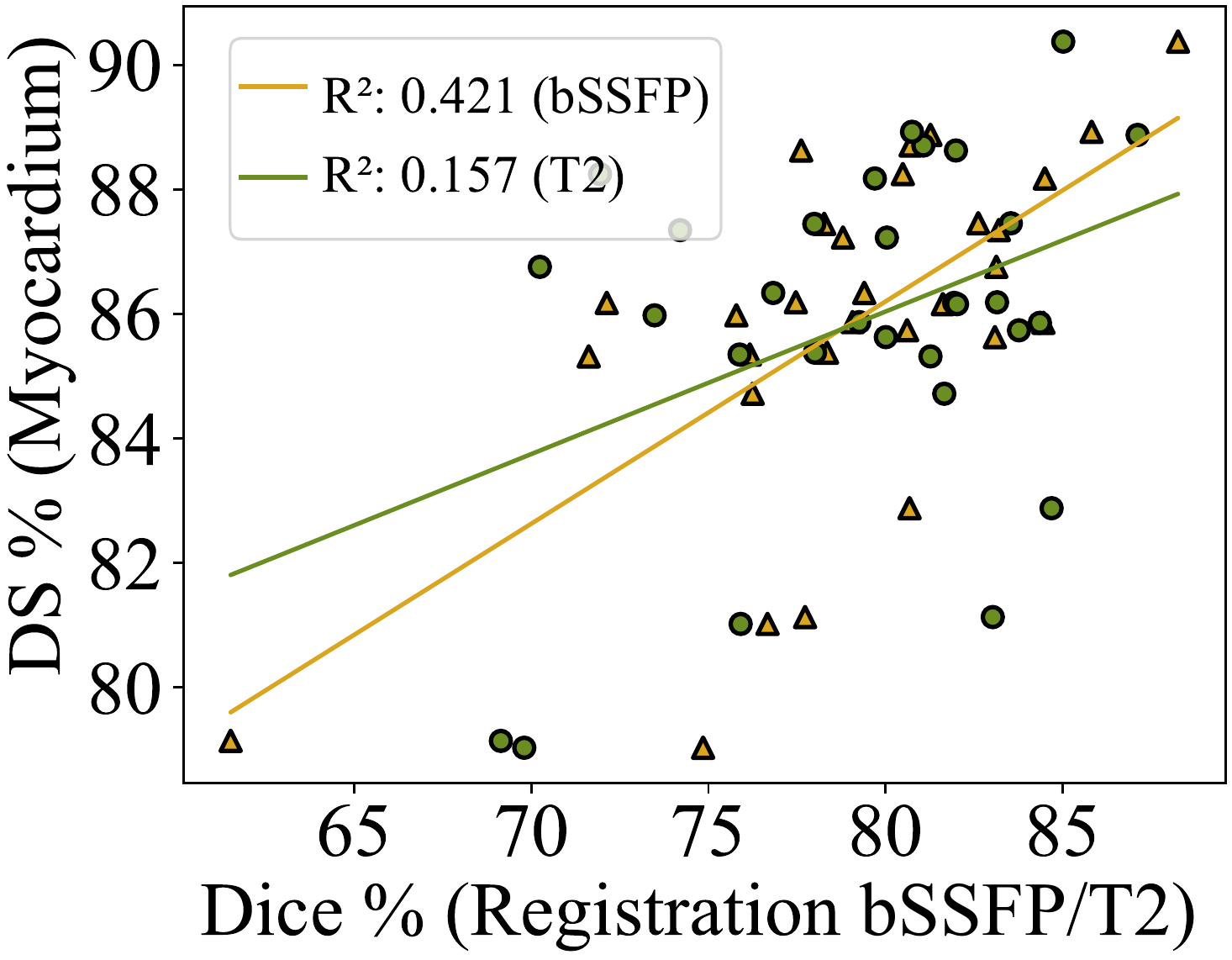}
			\caption{Correlation between the accuracy of  multi-sequence registration and  \hly{myocardium} extraction.}
			\label{fig:corr:reg_prior}
		\end{minipage}
	\end{figure}

	\subsubsection{Correlation between \hly{Myocardium} Extraction and MyoPS}\label{sec:corre:myo_myops}

	\MyFig \ref{fig:corr:prior_myops} shows the scar segmentation had a correlation with the accuracy of \hly{myocardium} extraction. This is reasonable as scars (in LGE images) can appear identical to surroundings (such as LV and RV blood pool) \cite{zhuang2018multivariate}. Thus \hly{myocardium} contours were critical to distinguish scars from surroundings. \hly{Besides}, the correlation between edema segmentation and \hly{myocardium} extraction was insignificant. 
	\hly{The potential reason is that U-MyoPS achieved sufficient myocardium extraction accuracy to promote edema segmentation.}
	Meanwhile, the edema (in T2 images) were more easily distinguishable from the surroundings (such as LV and RV blood pool). Thus, the accuracy of edema segmentation  would be less sensitive to the accuracy of \hly{myocardium}  extraction.

	\subsubsection{Correlation Between Multi-sequence Registration and \hly{Myocardium} Extraction} 
	
	\MyFig \ref{fig:corr:reg_prior} shows that both the registration accuracy of \hly{bSSFP} and T2 images were correlated to \hly{myocardium} extraction, while the registration accuracy of \hly{bSSFP} had more impact on \hly{myocardium} extraction than the one of T2. This was reasonable as \hly{bSSFP} images could provide clear boundary information, which was more important than T2 for \hly{myocardium} extraction.

	\subsection{Results on Public MYOPS2020 Challenge Dataset}\label{exp:results:MYOPS2020}
	
	We performed MyoPS for MYOPS2020 challenge dataset via U-MyoPS$_{\text{\hly{bLT}}}$. The gold standard pathology labels in the common space for evaluation were provided by MYOPS2020 challenge. Table \ref{tab:MYOPS2020} summarizes segmentation results of U-MyoPS$_{\text{\hly{bLT}}}$ as well as other state-of-the-art methods\footnote{http://www.sdspeople.fudan.edu.cn/zhuangxiahai/0/myops20/result.html}.  In this dataset, U-MyoPS$_{\text{\hly{bLT}}}$ achieved third and second places for scar and edema segmentation, respectively.  It is worth noting that U-MyoPS$_{\text{\hly{bLT}}}$ obtained segmentation results without using model ensemble strategies, whereas top performance methods (\ie UESTC*, UBA* and NPU*) achieved their results by using different model ensemble strategies  \cite{martin2020stacked,zhai2020myocardial,zhang2020efficientseg}. The lower part (under the dashed line) of Table \ref{tab:MYOPS2020} presents that U-MyoPS$_{\text{\hly{bLT}}}$ obtained the best average results among all the comparison methods when discarding model ensemble strategies.
	
\begin{figure}
		\centering     
		\subfigure[Predicted displacement of $R_{\hly{bSSFP}}$]{\label{fig:myops_tps_parameter:a}\includegraphics[width=0.49\textwidth]{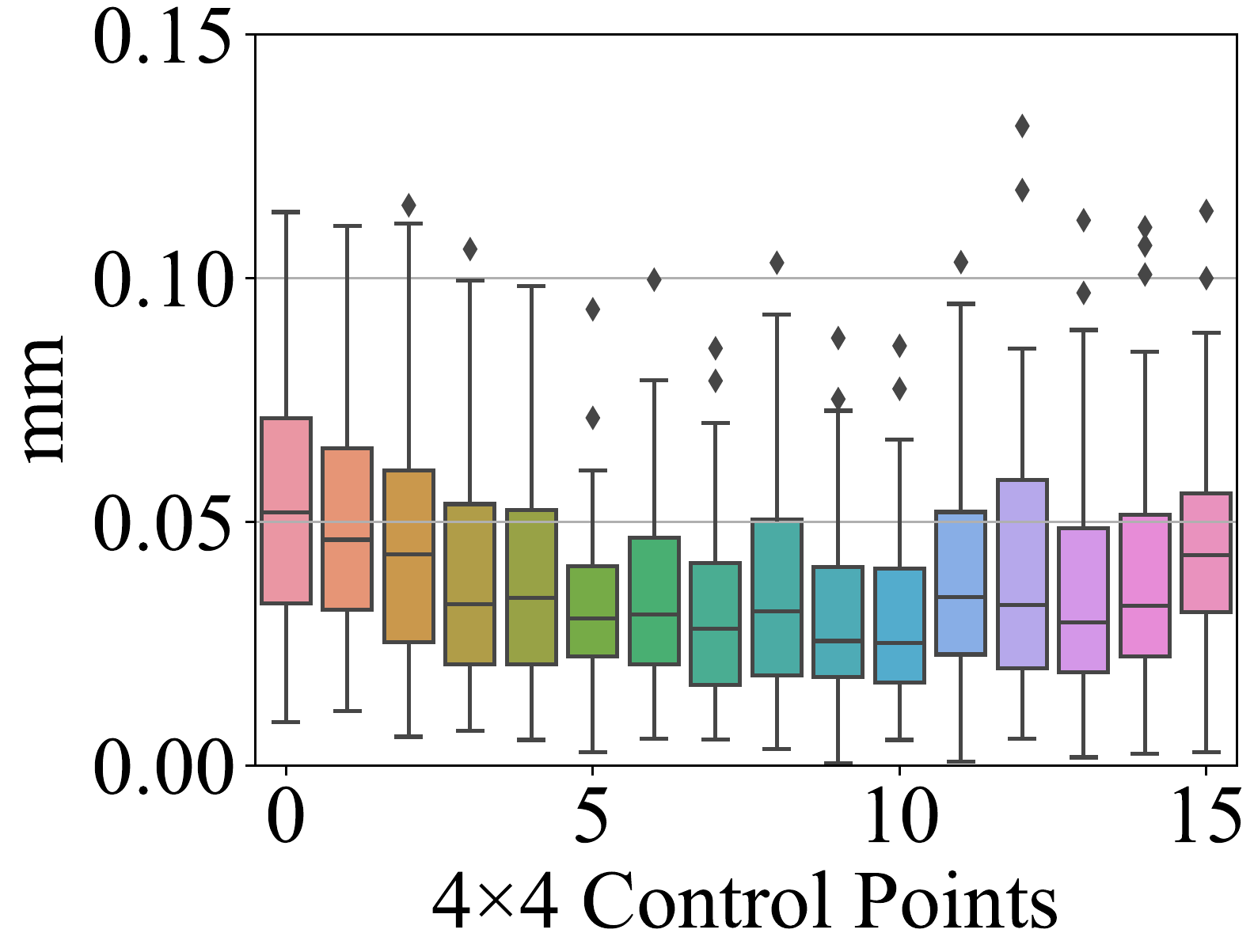}}
		\subfigure[Predicted displacement of $R_{T2}$]{\label{fig:myops_tps_parameter:b}\includegraphics[width=0.49\textwidth]{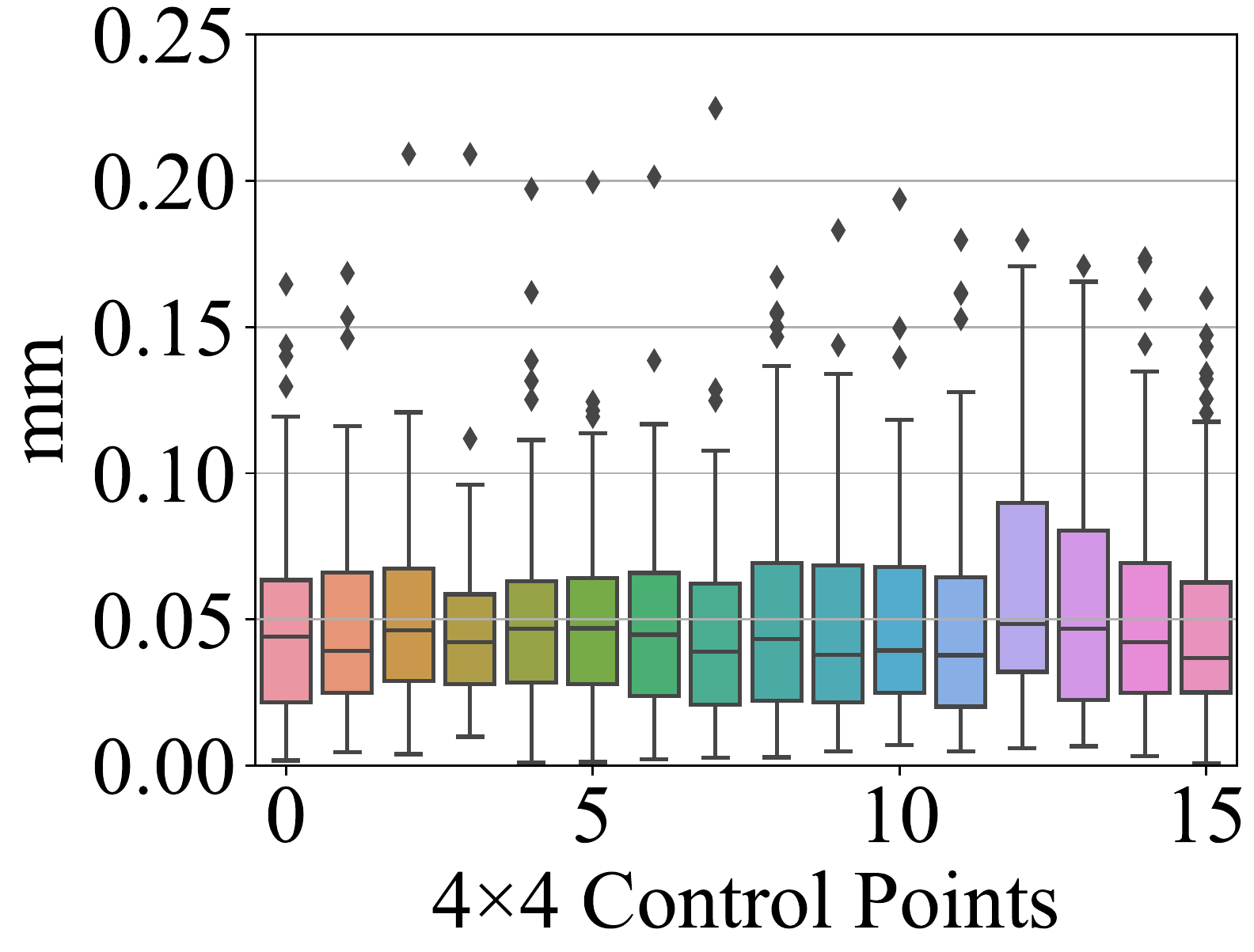}}
		\caption{Box plots of predicted TPS displacements on MYOPS2020 challenge dataset. For better demonstration, we normalized the displacements by dividing them by $H$ and $W$, and obtained the L2 norm of normalized displacements. (a) and (b) visualize the predicted displacements of $R_{\hly{bSSFP}}$ and $R_{T2}$, respectively. }
		\label{fig:myops_tps_parameter} 
	\end{figure}

\begin{figure}[h] 
		\centering
		\includegraphics[width=1.0\textwidth]{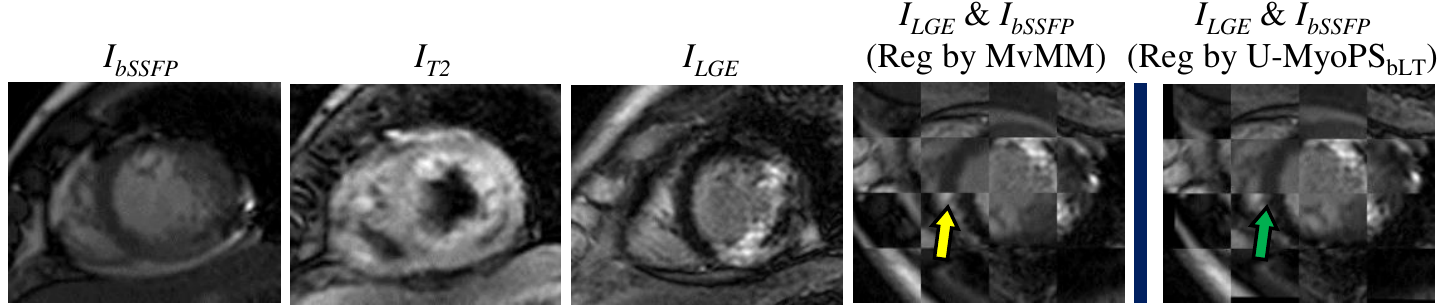}
		\caption{A checkerboard visualization of  multi-sequence cardiac magnetic resonance image registration result. The images are from \#19 case of test dataset of MYOPS2020 challenge. After registering by MvMM, there may remain misalignment between $I_{\hly{bSSFP}}$ and $I_{LGE}$ (see yellow Arrow). While, U-MyoPS$_\text{\hly{bLT}}$ could further improve the official registration results (see green Arrow). Reg: Registered.}
		\label{fig:unalinged_case}    
	\end{figure}

	
	Additionally, \MyFig \ref{fig:myops_tps_parameter} displays the statistic of predicted TPS displacements (see Eq. \ref{eq:tps_param}).  One can see most of the displacements were less than 0.05 mm.
	Such a phenomenon is reasonable since MYOPS2020 challenge dataset was officially pre-aligned.
	Nevertheless, a few displacements were larger than 0.2 mm. This is probably due to the fact that  some cases were not well aligned in the official MYOPS2020 challenge dataset \cite{li2022myops_new_version}.
	\hly{U-MyoPS$_\text{\hly{bLT}}$} could further alleviate the misalignment, as shown in \MyFig \ref{fig:unalinged_case}.

	\begin{table}[htb] 
		\resizebox{1\textwidth}{!}{
			\begin{tabular}{lccc}
				\hline
				Method	&  Scar (\%)  $\uparrow$ &   Edema (\%)  $\uparrow$& Average  (\%)  $\uparrow$\\
				\hline
				UESTC* \cite{zhai2020myocardial}  & 	 \textbf{67.2}   &  	\textbf{73.1}& \textbf{70.2}  \\
				
				UBA* \cite{martin2020stacked}  & 	 \underline{66.6}   &  	 69.8 & 68.2   \\
				NPU* \cite{zhang2020efficientseg} & 	64.7 &  	 70.9 & 67.8  \\
				
				\hdashline
				UESTC \cite{zhai2020myocardial}  & 	 {64.1}   &  	{69.5} &  66.8 \\
				NPU \cite{zhang2020efficientseg} & 	62.6  &  	 69.5 & 66.1 \\
				
				CQUPT II \cite{li2020cms} & 	58.1   &  	 72.5 & 65.3 \\
				
				U-MyoPS$_{\text{\hly{bLT}}}$ & 	{ 64.7 } 	 &  \underline{ 72.6 } &  \underline{68.6}\\
				
				\hline
				
			\end{tabular}
		}{
			\caption{The Dice results of different methods on the testing set of MYOPS2020 challenge dataset. Superscript ``*'' indicates that the method employed a model ensemble strategy. The best and second metrics are labeled in \textbf{bold} and \underline{underline}. \hly{The results of comparison methods were gathered from their official papers.} 
			}
			\label{tab:MYOPS2020}
		}
		
	\end{table}

	\section{Discussion and Conclusion}\label{sec:diss}

	This work presents an automatic \hly{combined computing} framework, \ie U-MyoPS, which could \hly{segment scar, edema and healthy myocardium within one unified result using practical unaligned MS-CMR images.}
	To our best knowledge,  existing \hly{DL-based} MyoPS methods  focused on processing aligned \hly{MS-CMR} images \cite{wang2022awsnet,li2022multi,li2022myops_new_version}. U-MyoPS could segment myocardial pathologies from unaligned \hly{MS-CMR} images in a fully automatic way, which could bring convenience to clinical practices. In U-MyoPS, we proposed a novel MSF for spatially unaligned feature map fusion. MSF could be considered as one of layer-level fusion schemes \cite{zhou2019review,ramachandram2017deep}. 
	In contrast to previous fusion strategies \cite{dolz2018hyperdense,zhou2020multi,jiang2020max,lv2019multi}, MSF provided a more appropriate way to aggregate spatially unaligned multi-sequence \hly{feature maps} (see \MyFig \ref{fig:rj:feat_MSF}). 
	Meanwhile, we introduced an SPG to propagate extracted \hly{myocardium} to promote MyoPS (see Section \ref{sec:exp:ablation:SPG}]). 
	In experiments, we evaluated U-MyoPS on an unaligned \hly{MS-CMR} dataset (see Table \ref{tab:rj:U-MyoPS}) and a public pre-aligned \hly{MS-CMR} dataset (see Table \ref{tab:MYOPS2020}). \hly{The results demonstrated that U-MyoPS achieved} comparable performance to existing MyoPS methods.

	The main limitation of U-MyoPS is the lack of joint optimization of multi-tasks. During our training period, we simultaneously optimized the multi-sequence registration and \hly{myocardium} extraction tasks, but independently optimized the pathology segmentation task. 
	Actually, we had performed joint optimization of these three tasks but observed a performance decrease of the pathology segmentation task.  This is because  both multi-sequence registration and \hly{myocardium} extraction tasks focused on structure information, so they  \hly{collaborated with} each other based on joint optimizations. Whereas the pathology segmentation task focused on the high-lighted abnormal intensity information, and thus it is still challenging to combine \hly{it} with multi-sequence registration and \hly{myocardium} extraction tasks at present. Future work could pursue an end-to-end trainable framework.

	Further improvements \hly{in} U-MyoPS could be achieved \hly{by} exploring the importance of each sequence.
	Note the effects of different sequences are varied. For example, aligning \hly{bSSFP} and T2 images promoted scar segmentation (see \MyFig \ref{fig:corr:reg_myops}). While the registration accuracy of \hly{bSSFP} brought more benefit to scar segmentation than the one of T2.  
	Thus, it is critical to estimate the reliability of each sequence for specific tasks. We can study dynamic methods for the weighted fusion of different images for a specific task in the future. \hly{Besides, there exist slice shifts due to breath-hold mis-registration in multi-slice CMR images, which could hinder myocardial pathology monitoring\cite {gilbert2019independent}. In future work, it is valuable to correct these shifts to promote U-MyoPS for clinical applications. }
	
	\hly{
		U-MyoPS is a new technique to perform myocardial pathology segmentation from unaligned \hly{MS-CMR} images. Aligning \hly{MS-CMR} images could promote  MyoPS performance.
		There exist three key points, \ie accuracy, time efficiency and generalizability, to consider in determining whether MyoPS algorithms ﬁt for the clinical workflow \cite{li2022myops_new_version}. For accuracy,  U-MyoPS could achieve agreements with manual delineation for scar size, scar transmurality and edema size quantification. For time efficiency, U-MyoPS consumed 10 seconds to segment pathologies from unaligned MS-CMR images, which was efficient for clinical applications. Nevertheless, for generalizability, U-MyoPS would suffer performance degradation when the test data acquisition protocol differs significantly from the training data acquisition protocol \cite{zhou2022domain,pei2021disentangle}.  
		Therefore, we need to further improve the generalization ability of U-MyoPS  before it could be applied for routine clinical assessment.  
	}

	\bibliographystyle{ieeetr}
	\bibliography{refs}
\end{document}